\documentclass[fleqn,usenatbib]{mnras}

\usepackage{newtxtext,newtxmath}
\usepackage[T1]{fontenc}

\DeclareRobustCommand{\VAN}[3]{#2}
\let\VANthebibliography\thebibliography
\def\thebibliography{\DeclareRobustCommand{\VAN}[3]{##3}\VANthebibliography}

\usepackage{graphicx}	% Including figure files
\usepackage{amsmath}	% Advanced maths commands
\usepackage{booktabs}   % Gives nicer formatting for Table 2
\usepackage{cuted}      % Used to display the large matrix in the appendix across two columns
\usepackage{upgreek}    % gives a nicer looking \mu symbol than the default, to represent "micro" as a prefix
\usepackage{hyperref}   % allows URL hyperlinks to publicly-available source code
\usepackage{xcolor}     % Allows colors for highlighting changes in response to reviewer comments
\usepackage{ulem}       % Allows strikethrough for highlighting changes in response to reviewer comments
\usepackage{braket}     % Allows <bra-ket> notation for cos(theta) terms
\usepackage{subcaption}     % Allows <bra-ket> notation for cos(theta) terms

\title[Aerosols are not Spherical Cows]{Aerosols are not Spherical Cows: Using Discrete Dipole Approximation to Model the Properties of Fractal Particles}

\author[M. G. Lodge]{
M. G. Lodge$^{1}$\thanks{E-mail: m.g.lodge@bristol.ac.uk},
H. R. Wakeford$^{1}$,
Z. M. Leinhardt$^{1}$
\\
$^{1}$School of Physics, HH Wills Physics Laboratory, Tyndall Avenue, Bristol BS8 1TL, UK\\
}

\date{Accepted 2023 November 28. Received 2023 November 8; in original form 2023 August 16.}

\pubyear{2023}

\begin{document}
\label{firstpage}
\pagerange{\pageref{firstpage}--\pageref{lastpage}}
\maketitle

\begin{abstract}
The optical properties of particulate-matter aerosols, within the context of exoplanet and brown dwarf atmospheres, are compared using three different models: Mie theory, Modified Mean Field (MMF) Theory, and Discrete Dipole Approximation (DDA). Previous results have demonstrated that fractal haze particles (MMF and DDA) absorb much less long-wavelength radiation than their spherical counterparts (Mie), however it is shown here that the opposite can also be true if a more varying refractive index profile is used. Additionally, it is demonstrated that absorption/scattering cross-sections, and the asymmetry parameter, are underestimated if Mie theory is used. Although DDA can be used to obtain more accurate results, it is known to be much more computationally intensive; to avoid this, the use of low-resolution aerosol models is explored, which could dramatically speed up the process of obtaining accurate computations of optical cross-sections within a certain parameter space. The validity of DDA is probed for wavelengths of interest for observations of aerosols within exoplanet and brown dwarf atmospheres ($0.2-15~\upmu$m). Finally, novel code is presented to compare the results of Mie, MMF and DDA theories (\texttt{CORAL}: Comparison Of Radiative AnaLyses), as well as to increase and decrease the resolution of DDA shape files accordingly (\texttt{SPHERIFY}). Both codes can be applied to a range of other interesting astrophysical environments in addition to exoplanet atmospheres, for example dust grains within protoplanetary disks.
%This is a simple template for authors to write new MNRAS papers.
%The abstract should briefly describe the aims, methods, and main results of the paper.
%It should be a single paragraph not more than 250 words (200 words for Letters).
%No references should appear in the abstract.
\end{abstract}

\begin{keywords}
Planets and satellites: Atmospheres -- Methods: Observational -- Radiative Transfer
\end{keywords}

%%%%%%%%%%%%%%%%%%%%%%%%%%%%%%%%%%%%%%%%%%%%%%%%%%

%%%%%%%%%%%%%%%%% BODY OF PAPER %%%%%%%%%%%%%%%%%%

\section{Introduction}

Transmission, reflection, and emission spectra have allowed us to unlock a wide variety of interesting information about exoplanet and brown dwarf atmosphere compositions, pressure and temperature structures, scale heights and other interesting climatological data. Advanced theoretical models are required to interpret the spectra, and extensive efforts have been made to develop retrieval algorithms, forward models and general circulation models to probe and understand the physical and chemical processes at play. All of these fundamentally require the interaction between light and particles in their atmospheres --  defined in terms of their optical (extinction, scattering and absorption) cross-sections -- to be modelled as accurately as possible. Aerosols (in the form of particulate matter within the atmosphere) have been shown to have an important role in the optical transmission of light above the planet's surface, as well as the feedback mechanisms of the climate as a whole \citep[see review by][]{gao2021aerosols}. The diversity of the potential chemical compositions of aerosols, as well as their formation mechanisms and physical structures, has led to several studies that have attempted to determine their impact. Some studies have been dedicated to exploring the effects of the particle shapes \citep{arney2017pale} or composition/size distribution \citep{wakeford2015transmission}, which follow from the modelling of clouds and the micro-physical processes that are predicted to take place in their respective environments \citep{rooney2021modeling, turco1979one, helling2008consistent}. As a direct consequence of the formation and evolutionary processes considered, many different particle shapes and structures are predicted to form, which can make the task of modelling them overwhelming; however, we can gain insights into their physical properties by analysing real aerosols that occur in the Earth's atmosphere.
		
\begin{figure}
	\includegraphics[width=\columnwidth]{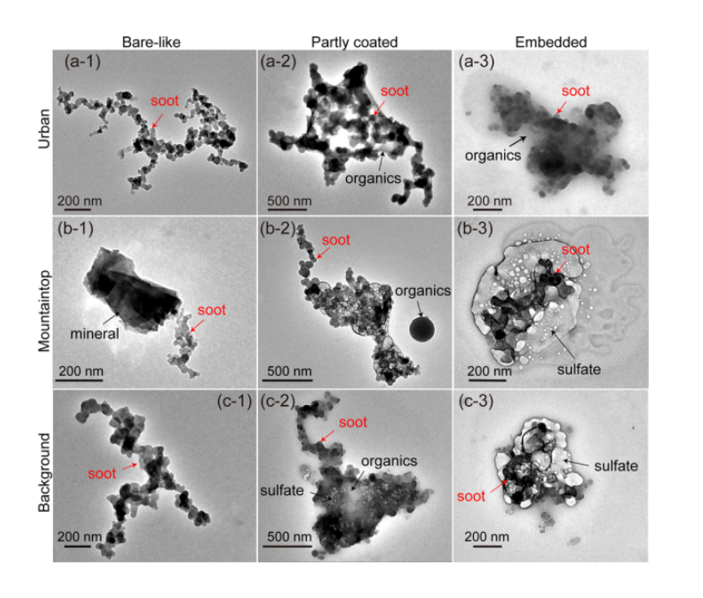}
    \caption{TEM images of individual soot particles classified into three types: bar-like (a1-c1), partly coated (a2-c2), and embedded (a3-c3), reprinted (adapted) with permission from \citet{wang2017fractal}. Copyright {2017} American Chemical Society.}
    \label{fig:TEM_fractals}
\end{figure}

To reduce the number of free parameters in these models, and to avoid the expensive computation time required to process large numbers of interactions and particle sizes at high-resolution, studies often consider interactions in the Mie region (where the wavelength is roughly equal to the radius of the particle) to occur from homogenous spheres \citep{benneke2019sub, lacy2020jwst, lee2014constraining, zhang2019forward, samra2022mineral}. In this regime, interference effects begin to dominate, so simple geometric optics/Rayleigh scattering theories cannot be used. However, it is recognised that the assumption of spherical particles may be a simplification. Aerosols on Earth often take the form of more complex, fractal shapes, as observed in the tunnelling electron microscope (TEM) images in Figure~\ref{fig:TEM_fractals} \citep{wang2017fractal}. Alternatively, they can form into approximately spherical shapes as they grow large or mix with condensates. Depending on which shape they form, the optical properties of these aerosols can be very different.

\citet{marley2013clouds} highlight the importance and general challenges of cloud modelling with non-spherical particles. As a local analogy, they describe how single-scattering properties of ice particles in the terrestrial atmosphere varies with surface roughness and shape (which itself can depend on temperature). \citet{samra2022mineral} highlight the importance of these geometrical considerations, finding that retrievals of exoplanets and brown dwarfs are often biased by the assumption of compact spherical particles, and they explore the extent to which non-spherical particles increase the opacity of silicate spectral features. There has also been significant interest in describing the transmission spectra of GJ 1214b using fractal aerosols in cloud models; \citet{adams2019aggregate} found that hydrocarbon aggregates were much more successful at reproducing the spectra than their spherical counterparts, and \citet{ohno2020clouds} performed a detailed study using size distributions of porous non-spherical KCl particles. \citet{ohno2021grain} also examined the effects of porous, fractal aerosols in cloud models of super-puffs.

The geometric shape of the atmospheric particles can actually effect how light is scattered in the atmosphere so significantly that it could impact the energy balance of the entire planet \citep{schuerman1980light, bohren2008absorption}. \citet{wolf2010fractal} showed that the fractal nature of particles in hazes on Titan and the Archean Earth may resolve the faint young sun paradox. This paradox arises because stellar models predict that during the first billion years of the Earth's history, the Sun would have provided less than 70 \% of today's energy input to the Earth's atmosphere. However, the Earth is thought to have still been warm during much of this period, because there is no evidence of widespread glaciation \citep{sagan1972earth}. In addition, the early life forms that are believed to exist would have struggled to survive the intense UV radiation that is predicted to be emitted from the young Sun. \citet{wolf2010fractal} modelled haze particles as fractal aggregates instead of spherical particles and this was shown to increase transparency to longer wavelengths of light (allowing the Earth to heat up). Additionally, using a fractal aerosol model dramatically increased absorption of shorter wavelengths, protecting early life from the higher flux of UV radiation, potentially solving the faint young sun paradox. Here we revisit their analysis with updated models and for a different refractive index profile.

Atmospheric models, retrievals, and radiative transfer codes all require accurate extinction, absorption and scattering cross-sections to produce reliable outputs (see \citet{des2008rayleigh} for a typical example of their use). Although Mie spheres are used regularly, many attempts have been made to model aerosols more realistically, ranging from simple analytical theories to complex and exact numerical simulations. Effective medium theory estimates the optical properties of multi-composite aerosols by averaging the refractive index of each material involved \citep{maxwell1904xii}. Rayleigh-Debye-Gans theory has been shown to be able to have some applicability to fractal aerosols that are ``optically soft'', which is where $|m-1|<<1$ if $m$ is the complex refractive index \citep{wang2002experimental}. Separately, exact analytical solutions have been determined for simple shapes other than spheres, such as homogeneous cylinders and ellipsoids \citep{bohren2008absorption}. The optical properties of complex particles can also be simulated by modelling them as distributions of hollow spheres \citep{min2005modeling}, which has had some success in representing ensembles of particles. A rigorous solution for multi-sphere (and multi-composite) structures was developed by \citet{xu1995electromagnetic}, however this is only valid for isotropic, spherical components. Modified Mean Field (MMF) theory \citep{berry1986optics, botet1997mean, Tazaki_Tanaka_2018} simplifies aggregate shapes by defining them in terms of a few simple fractal parameters, and reproduces cross-sections very well in some cases, but all monomers (composite spheres that make up the aggregate) are assumed to be the same material and size. All of the above models are useful in certain contexts, but have some limitations, or make assumptions/simplifications that mean that the cross-sections obtained may not be correct in all cases. Finally, very flexible numerical treatments, such as Discrete Dipole Approximation (DDA) \citep{draine1994discrete} and T-matrix \citep{waterman1965matrix, mishchenko1996t} were developed to deal with any shape, structure and material type comprehensively, and these final two approaches are traditionally used to provide a benchmark for the accuracy of other models. Out of the two, DDA was chosen as a benchmark model for this paper due to it's flexibility in dealing with multi-composite materials. 

Each model has it's own advantages and disadvantages, but as a general rule, more accurate and complex models come with a cost of significantly increased computation time. Determining a fast and accurate method to calculate optical properties would be of significant interest, especially in a field where computations are already pushed to the limit.

This study considers three such models; Mie theory \citep{bohren2008absorption}, Modified Mean Field (MMF) theory \citep{berry1986optics, botet1997mean, Tazaki_Tanaka_2018}, and Discrete Dipole Approximation (DDA - first introduced by \cite{purcell1973scattering}, and developed further for the analysis of interstellar dust grains by \citealt{draine1994discrete}). We explore each of these in more depth below, but to summarize - Mie is the simplest and quickest (assuming spherical particles), MMF gives a better approximation of particle shapes but the values obtained are not accurate in all cases, and DDA is the most accurate but the slowest. Each theory has distinct advantages and disadvantages; Mie gives an exact analytical solution for perfect spheres, but it can not compute accurate particle cross-sections for particles of more complex geometry. MMF characterises aerosol particles using a few simple fractal structure parameters (e.g. fractal dimension); it is fast, and can quickly compute branched structures with large numbers of particles connected in chains without increasing computation time. However the monomers have to be the same material type and size, and cross-sections (for absorption in particular) can be inaccurate by up to 40\%. The most accurate theory studied, DDA, can calculate cross-sections for aerosols composed of any shape and material composition, which makes it very powerful and a widely used theory in aerosol science; however, it is often assumed to require high levels of computing power, which would make it unfeasible to use in atmospheric retrievals. This assumption is explored in this paper.

In this paper we compare Mie, MMF and DDA for a range of realistic particle shapes and structures, for the specific range of aerosol sizes and types that we might expect to find in exoplanet and brown dwarf atmospheres. We also aim to provide a pathway for other researchers to model the scattering and absorption cross-sections accurately within these regions (with as small a computational cost as possible), using the new publicly available code \texttt{CORAL}: Comparison Of Radiative AnaLyses (see Section \ref{sec:computational_methods}). Cross-sections can then be calculated for radiative transfer codes (for example) either in-situ, or as a pre-calculated grid, in the fastest and most accurate way. 

In section \ref{sec:Theoretical_Basis} we discuss the mathematical basis for each of the three theories that we wish to compare: Mie, MMF and DDA. In section \ref{sec:methods} we discuss how the realistic aerosol shapes were obtained, how fair inter-comparisons between the three models were made, and include technical details of a new code, \texttt{SPHERIFY}, which can be used to increase or decrease the resolution (number of dipoles) of the aerosol particles used for DDA. In section \ref{sec:results} we discuss the results of the comparison for three realistic aerosol shapes, including revisiting the results of \cite{wolf2010fractal} with a new refractive index profile, and in section \ref{sec:conclusions} we summarise conclusions and our recommendations for modelling aerosols in astrophysical atmospheres going forwards.

\section{Theoretical Basis} \label{sec:Theoretical_Basis}

\begin{figure}
	\includegraphics[width=\columnwidth]{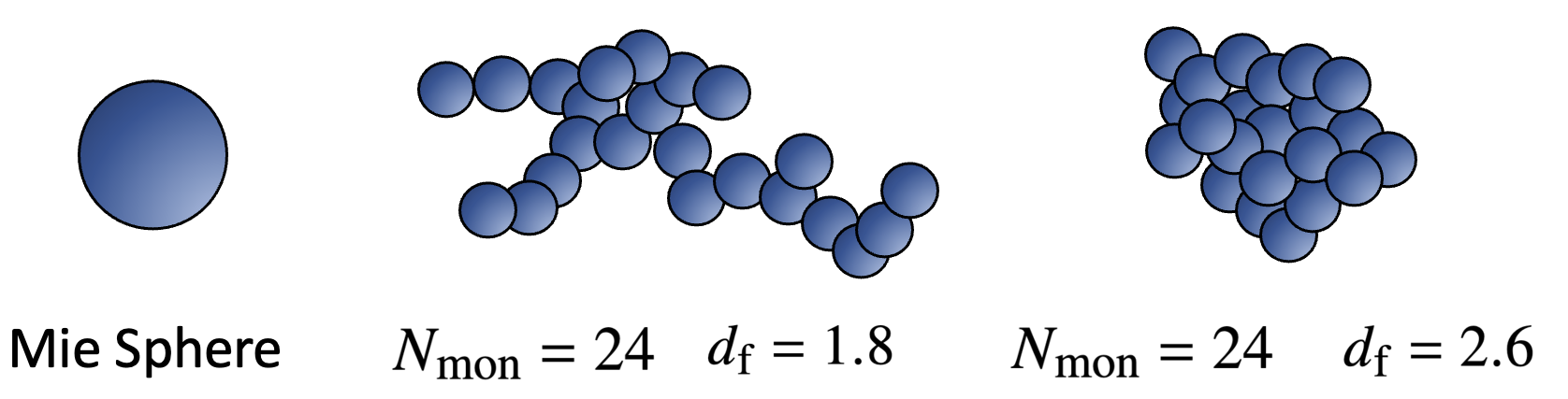}
    \caption{The material in the three aerosols shown above has the exact same volume in each case, but they would have very different scattering and absorptive properties due to their geometries. Fractal aggregates can be characterised by their number of monomers ($N_\mathrm{mon}$) and fractal dimension ($d_{f}$) -- as a general rule, aggregates that are more compact have higher fractal dimensions.}
    \label{fig:sphere_vs_fractal}
\end{figure}

To compare the amount of light absorbed and scattered from fractal aerosols versus a homogeneous sphere of the same mass (Figure~\ref{fig:sphere_vs_fractal}), the key optical parameters that we need to obtain are $C_\mathrm{sca}$ and $C_\mathrm{abs}$, the total cross-sectional area of a beam of incident radiation that is scattered and absorbed by a particle respectively, and $C_\mathrm{ext}$, the extinction cross-section (which represents the total area ``removed'' from the beam, equal to the sum of $C_\mathrm{sca}$ and $C_\mathrm{abs}$). In addition, the asymmetry parameter $g$ (the proportion of light that is scattered forwards or backwards) is an important quantity for haze calculations in solar system planets \citep{lavvas2010titan, rannou1997new} and exoplanets \citep{garcia2018exoplanet, robinson2017theory}. The mathematical process of determining each of these parameters for each of Mie, MMF and DDA theories are described below.

\subsection{Mie Theory - Spheres}

The exact solution for calculating  Mie scattering from homogeneous spheres is well known. Here we briefly summarise the key equations, but see \citet{bohren2008absorption} for a derivation of the theory, beginning with the fundamental interaction of electromagnetic radiation and Maxwell's equations. To determine the optical cross-section of a spherical particle, we first obtain the Mie coefficients:
\begin{equation}
    a_{n}=\frac{\mu m^{2}j_{n}(mx)[xj_{n}(x)]'-\mu_{1}j_{n}(x)[mxj_{n}(mx)]'}{\mu m^{2}j_{n}(mx)[xh_{n}^{(1)}(x)]'-\mu_{1}h_{n}^{(1)}(x)[mxj_{n}(mx)]'},
\end{equation}
\begin{equation}
    b_{n}=\frac{\mu_{1}j_{n}(mx)[xj_{n}(x)]'-\mu j_{n}(x)[mxj_{n}(mx)]'}{\mu_{1}j_{n}(mx)[xh_{n}^{(1)}(x)]'-\mu h_{n}^{(1)}(x)[mxj_{n}(mx)]'},
\end{equation}
where $m$ is the complex refractive index of the sphere material, $x=\frac{2\pi r}{\lambda}$ is the size parameter for a sphere of radius $r$ and wavelength $\lambda$, the terms $\mu$ and $\mu_{1}$ are the magnetic permeability of the ambient medium and sphere material respectively. The terms $j_{n}(z)$ and and $h_{n}^{(1)}(z)$ represent the spherical Bessel and Hankel functions respectively (both 1st kind) for their given arguments (either $x$ or $mx$). The primed terms represent derivatives with respect to their argument, calculated using the relationship $[zj_{n}(z)]'=j_{n-1}(z)-nj_{n}(z)$.

Once the Mie coefficients have been obtained for a given sphere size and wavelength, the following formulae are used to calculate the extinction, scattering and absorption cross-sections:	
\begin{equation} \label{eq:C_sca_mie}
    C_\mathrm{sca}=\frac{2\pi}{k^{2}} \sum_{n=1}^{\infty}(2n+1)(\lvert a_{n}\rvert^{2} + \lvert b_{n}\rvert^{2}),
\end{equation}

\begin{equation} \label{eq:C_ext_mie}
    C_\mathrm{ext}=\frac{2\pi}{k^{2}} \sum_{n=1}^{\infty}(2n+1) \mathrm{Re}(a_{n}+b_{n}),
\end{equation}

\begin{equation} \label{eq:C_abs_mie}
    C_\mathrm{abs}=C_\mathrm{ext}-C_\mathrm{sca}.
\end{equation}
Typically, because the terms decrease in size as the sum progresses, we follow the method outlined by \citet{bohren2008absorption} to truncate the infinite series in Eqs.~\ref{eq:C_sca_mie} and \ref{eq:C_ext_mie} at:
\begin{equation} \label{eq:truncation}
    n_\mathrm{max}=x+4x^\frac{1}{3}+2.
\end{equation}

We also commonly define $Q_\mathrm{ext}=\frac{C_\mathrm{ext}}{\pi r^{2}}$ (and similar for $Q_\mathrm{sca}$ and $Q_\mathrm{abs}$) to represent extinction ``efficiency''. To put this in physical terms, a value of 1 would mean the sphere ``removed'' an area of radiation equal to its geometrical cross-section. However, the interesting properties of waves allow values that are counter-intuitive and often much larger than this, because of the interference and diffraction effects that occur when $r \approx \lambda$ (see ``Extinction Paradox'' for further reading and proof in \citealt{bohren2008absorption}).

Finally, the asymmetry parameter $g=\braket{\cos \theta}$ can be calculated. This is a single-value introduced by \cite{henyey1941diffuse} to characterise the average direction of scattered light. The value ranges from $-1 \leq g \leq 1$, representing a spectrum of fully back-scattered ($g=-1$), isotropic ($g=0$), and fully forward-scattered radiation ($g=1$). It can be calculated for spherical particles using:
\begin{multline}
    g = \frac{4}{x^2 Q_\mathrm{sca}} \Biggl( \sum_{n=1}^{\infty} \frac{n(n+2)}{n+1} \mathrm{Re} \left( a_{n}a_{n+1}^* +  b_{n}b _{n+1}^{*} \right) \\ 
    + \sum_{n=1}^{\infty} \frac{2n+1}{n(n+1)} \mathrm{Re} \left( a_{n}b_{n}^{*} \right) \Biggr).
\end{multline}

\subsection{Modified Mean Field Theory (MMF) - Fractal Aggregates}

Mean Field Theory (MFT) was originally formulated by \citet{berry1986optics} and further developed by \citet{botet1997mean} to better describe the optical properties of fractal aggregates, improving on the values obtained by Mie theory. \citet{Tazaki_Tanaka_2018} demonstrated that although MFT reproduces extinction cross-sections well, it did not always result in accurate absorption cross-sections. To rectify this, they introduced Modified Mean Field Theory (MMF), which adds a final modifying term to the MFT cross-sections (based on the geometrical cross-section of the aggregate) to attain a better approximation of absorption and scattering cross-sections.

To calculate these cross-section, we begin by defining our fractal aggregate as a collection of $N_\mathrm{mon}$ identical spheres (monomers) of radius $R_0$. We can define:
\begin{equation} \label{fractal_equation}
    N_\mathrm{mon}=k_{0} \left(\frac{R_{g}}{{R_0}}\right)^{d_{f}},
\end{equation}
where $d_{f}$ is the fractal dimension (roughly, a description of the shape type - see Figure~\ref{fig:sphere_vs_fractal}), $k_0$ is the fractal prefactor (a scaling relation), and $R_{g}$ is the radius of gyration of the aggregate (the radial distance to a point which would have a moment of inertia the same as the body's actual distribution of mass, if the total mass of the body were concentrated there - see section \ref{section:comparisons_between_models}). We can also define a size parameter for the aggregates: $x_{g}=\frac{2\pi R_{g}}{\lambda}$.

Once we have defined a shape, the fundamental equation for calculating the extinction cross-section for a collection of $N_\mathrm{mon}$ spheres has a very similar form to Eq.~\ref{eq:C_ext_mie}:
\begin{equation} \label{C_ext_MMF}
    C_\mathrm{ext}^{\mathrm{MFT}}=\frac{2\pi N_\mathrm{mon}}{k^{2}} \sum_{n=1}^{\infty}(2n+1) \mathrm{Re}\left( \bar{d}_{1,n}^{(1)}+\bar{d}_{1,n}^{(2)} \right),
\end{equation}	
where $\bar{d}_{1,n}^{(1)}$ and $\bar{d}_{1,n}^{(2)}$ are the coefficients of the mean field, found from a set of linear equations defined by:

\begin{equation} 
    \bar{d}_{1,n}^{(1)} = a_{n} \left( 1 - (N_\mathrm{mon}-1) \sum_{\nu=1}^{\infty} \bar{A}_{1,n}^{1,\nu} \bar{d}_{1,\nu}^{(1)} + \bar{B}_{1,n}^{1,\nu} \bar{d}_{1,\nu}^{(2)} \right),
\end{equation}	

\begin{equation}
    \bar{d}_{1,n}^{(2)} = b_{n} \left( 1 - (N_\mathrm{mon}-1) \sum_{\nu=1}^{\infty} \bar{B}_{1,n}^{1,\nu} \bar{d}_{1,\nu}^{(1)} + \bar{A}_{1,n}^{1,\nu} \bar{d}_{1,\nu}^{(2)} \right).
\end{equation}
Coefficients $\bar{A}_{1,n}^{1,\nu}$ and $\bar{A}_{1,n}^{1,\nu}$ are calculated using:
\begin{multline}
    \bar{A}_{1,n}^{1,\nu} = \frac{2\nu +1}{n(n+1)\nu (\nu +1)} \sum_{p=|n-\nu|}^{n+\nu} [n(n+1)+\nu (\nu+1) - (p(p+1)] \\ a(\nu,n,p)s_{p}(kR_{g}),
\end{multline}

\begin{multline}
    \bar{B}_{1,n}^{1,\nu} = 2\frac{2\nu +1}{n(n+1)\nu (\nu +1)} \sum_{p=|n-\nu|}^{n+\nu} [n(n+1)+\nu (\nu+1) - (p(p+1)] \\ b(\nu,n,p)s_{p}(kR_{g}),
\end{multline}
where $a(\nu,n,p)$ and $b(\nu,n,p)$ are given by the integrated combinations of Legendre Polynomials $P_{l}(x)$ and associated Legendre Polynomials $P_{l}^{m}(x)$ below, and where $s_{p}(kR_{g})$ is a factor that depends on the structure of the dust aggregate: 

\begin{equation}
    a(\nu,n,p)= \frac{2p+1}{2} \int_{-1}^{1} P_{v}^{1}(x) P_{n}^{1}(x) P_{p}(x) dx,
\end{equation}

\begin{equation}
    b(\nu,n,p)= \frac{2p+1}{2} \int_{-1}^{1} P_{v}^{1}(x) P_{n}^{1}(x) \frac{dP_{p}(x)}{dx} dx,
\end{equation}

\begin{equation} \label{structure_term_MMF}
    s_{p}(kR_{g})= \frac{1}{2x_{g}^{d_{f}}} \int_{u_{\mathrm{min}}}^{u_{\mathrm{max}}} u^{d_{f}-1} j_{p}(u) h_{p}^{(1)}(u) f_{c} \left(\frac{u}{x_{g}}\right) du.
\end{equation}
In Eq.~\ref{structure_term_MMF}, $j_{p}(u)$ and $h_{p}^{(1)}$ are the spherical Bessel functions and Hankel functions of the 1st kind (respectively), and $f_{c} \left(\frac{u}{x_{g}}\right)$ is the cut-off function. There is some debate over the optimum equation to use here, but we choose the same formulation as in \citet{Tazaki_Tanaka_2018} and \citet{botet1995sensitivity} (the ``fractal dimension cutoff model''), given by:
\begin{equation}  \label{cutoff_function}
    f_{c} \left(\frac{u}{x_{g}}\right) = \frac{d_{f}}{2}\exp{\left( -\frac{1}{2} \left(\frac{u}{x_{g}} \right)^{d_{f}} \right)}.
\end{equation}
Equations \ref{C_ext_MMF} to \ref{cutoff_function} can be combined to calculate $C_\mathrm{ext}^{\mathrm{MFT}}$, and to calculate $C_\mathrm{sca}^{\mathrm{MFT}}$ we use:
\begin{equation} \label{C_sca_MFT}
    C_\mathrm{sca}^{\mathrm{MFT}}=\frac{2\pi}{k^{2}} \int_{0}^{\pi} S_\mathrm{11,agg} \sin{\theta} d\theta,
\end{equation}
where the scattering coefficient is given by:
\begin{equation} \label{S_11,agg}
    S_\mathrm{11,agg} = N_\mathrm{mon} S_{11}^{0} \left(1+(N_\mathrm{mon}-1)S(q) \right).
\end{equation}
Here $q=2k\sin{\left(\frac{\theta}{2}\right)}$, and $S(q)$ is the structure factor, given by:
\begin{equation}
    S(q) = \frac{c d_{f}}{q R_{g}} \int_{0}^{x_{\mathrm{max}}} x^{d_{f}-2} \sin{(q R_{g} x)} \exp{(-c x^{d_{f}})} dx.
\end{equation}
For this study, we use $c=0.5$ and $x_{\mathrm{max}}=50^{1/d_{f}}$, as in \citet{Tazaki_Tanaka_2018}. $S_{11}^{0}$ in Eq.~\ref{S_11,agg} is given by combining the following sets of equations:

\begin{equation}
    S_{11}^{0} = \frac{1}{2} \left( |S_{1}^{0}|^{2} + |S_{2}^{0}|^{2} \right),
\end{equation}

\begin{equation} \label{eq:S_1}
    S_{1}^{0} = \sum_{n=1}^{\infty} \frac{2n+1}{n(n+1)} \left( \bar{d}_{1,n}^{(1)} \pi_{n} + \bar{d}_{1,n}^{(2)} \tau_{n} \right),
\end{equation}

\begin{equation} \label{eq:S_2}
    S_{2}^{0} = \sum_{n=1}^{\infty} \frac{2n+1}{n(n+1)} \left( \bar{d}_{1,n}^{(1)} \tau_{n} + \bar{d}_{1,n}^{(2)} \pi_{n} \right).
\end{equation}
The scattering amplitudes in Eq.~\ref{eq:S_1} and \ref{eq:S_2}, are found by computing $\pi_{n} = \frac{P_{n}^{1}}{\sin{\theta}}$ and $\tau_{n} = \frac{dP_{n}^{1}}{d\theta}$, where $P_{n}^{1}$ are associated Legendre Polynomials. Finally, we can use Eq.~\ref{C_ext_MMF} and \ref{C_sca_MFT} to calculate $C_\mathrm{abs}^\mathrm{MFT}$:
\begin{equation} \label{C_abs_MFT}
    C_\mathrm{abs}^{\mathrm{MFT}}= C_\mathrm{ext}^{\mathrm{MFT}} - C_\mathrm{sca}^{\mathrm{MFT}}.
\end{equation}
We can then calculate the asymmetry parameter using:
\begin{equation}  \label{g_MMF}
    g = 2\pi \int_{0}^{\pi} \frac{S_{11,\mathrm{agg}}}{C_\mathrm{sca}k^{2}} \sin\theta \cos\theta \mathrm{d}\theta,
\end{equation}
The method above is used to determine the cross-sections used in \citet{wolf2010fractal}, however it is worth noting that while the extinction cross-sections are expected to be relatively accurate, \citet{Tazaki_Tanaka_2018} reports that MFT can significantly underestimate/overestimate $C_\mathrm{abs}$ and $C_\mathrm{sca}$. To improve these estimates, we can  ``modify'' these two terms (hence ``Modified''
Mean Field Theory): the accuracy can be greatly improved by adjusting the absorption cross-section value by a factor related to the Geometrical cross-section of the shape (see Appendix \ref{appendix:geo_cross_section} for details, based on method of \citet{Tazaki_2021}), and then taking whichever is larger from:

\begin{equation}  \label{C_abs_MMF}
    C_\mathrm{abs}^\mathrm{MMF} = \max{\left(C_\mathrm{abs}^{\mathrm{MFT}},G(1-\exp{(-\tau)})\right)},
\end{equation}
where $\tau$ is given by:

\begin{equation} 
    \tau = \frac{C_\mathrm{abs}^\mathrm{RDG}}{G}.
\end{equation}
$C_\mathrm{abs}^\mathrm{RDG}$ is the absorption cross-section obtained from Rayleigh-Debye-Gans (RDG) theory, given by:

\begin{multline} \label{C_abs_RDG}
    C_\mathrm{abs}^{\mathrm{RDG}}=\frac{2\pi N_\mathrm{mon}}{k^{2}} \sum_{n=1}^{\infty}(2n+1) \mathrm{Re} \Biggl( |a_{n}|^2 \left( \frac{1}{a_{n}*} - 1 \right) \\ + |b_{n}|^2 \left( \frac{1}{b_{n}*} - 1 \right) \biggl),
\end{multline}
where $*$ denotes complex conjugates. Finally, we can use the modified $C_\mathrm{abs}^{\mathrm{MMF}}$ term from Eq.~\ref{C_abs_MMF} and the unchanged extinction cross-section (that is, $C_\mathrm{ext}^{\mathrm{MMF}}=C_\mathrm{ext}^{\mathrm{MFT}}$) to calculate the scattering cross-section:

\begin{equation}
    C_\mathrm{sca}^{\mathrm{MMF}}= C_\mathrm{ext}^{\mathrm{MMF}} - C_\mathrm{abs}^{\mathrm{MMF}}.
\end{equation}

MMF offers great improvement to the accuracy of the optical cross-sections obtained over a wide range of wavelengths, improving the values obtained by MFT (from $\approx90\%$ error to around $30-40\%$ error at worst, but often much better). Additionally, this theory has a distinct advantage in that increasing the number of monomers $N_\mathrm{mon}$ does not increase computation time, allowing fast numerical study of large structures composed of many distinct monomers arranged in complex shapes. However, there are other light-scattering theories (for example DDA, discussed below) that offer even more impressive accuracy, but usually at the cost of increased computation time.

\subsection{Discrete Dipole Approximation (DDA) - Fractal Aggregates} \label{Theory:DDA}

The fundamental theory of DDA is based on the Clausius-Mossetti prescription, where it was determined that the bulk optical properties of material could be explained by a continuum of smaller particles (cubic dipoles - see Figure~\ref{fig:realistic_fractals}) with individual polarisabilities $\alpha_{j}$. The theory is exact when the number of dipoles $N\rightarrow\infty$, but in practise becomes an approximation limited only by computing power available and programming efficiency. 

To determine the optical cross-sections for a particle, we begin with an incident electric field vector defined by:

\begin{equation} \label{Eq:E_inc}
    \mathbf{E_{inc}}  =\mathbf{E_{0}}\exp(i\mathbf{k.r}) = \mathbf{\hat{e_{r}}}\exp{\left(i\frac{2\pi}{\lambda} \mathbf{\hat{k}.r} \right) },
\end{equation}	
where $\mathbf{\hat{k}}$ is the incident light vector after being rotated by polar angles $\theta$ and $\phi$: that is, for a right-handed rotation: $\mathbf{\hat{k}}=(\cos{\phi}\cos{\theta}, -\sin{\phi}\cos{\theta}, -\sin{\theta})$. Vector $\mathbf{r}$ is the absolute relative position of the electric field and a dipole: $\mathbf{r}=(x,y,z)$. Vector $\mathbf{\hat{e_{r}}}$ is the Cartesian components of the unit vector of the initial polarisation state of the field $(p_{x}, p_{y}, p_{z})$ after being rotated by polar angles $\theta$ and $\phi$:

\begin{equation} \label{polarisation_state_of_EM_field}
  \mathbf{\hat{e_{r}}}=
  \begin{pmatrix}
    \cos{\phi}(p_{x}\cos{\theta} + p_{z}\sin{\theta}) + p_{y}\sin{\phi} \\
    -\sin{\phi}(p_{x}\cos{\theta} + p_{z}\sin{\theta}) + p_{y}\cos{\phi} \\
    -p_{x}\sin{\theta} + p_{z}\cos{\theta}
  \end{pmatrix}.
\end{equation}

These rotations allow us to consider light incident from any direction in 3D space. When this field is incident on an aerosol composed of $N$ dipoles the problem involves solving the following series (compact form) of $3N$ linear equations:

\begin{equation} \label{compact_linear}
    \sum_{k=1}^{N}\mathbf{A_{jk}P_{k}}=\mathbf{E_{inc,j}}
\end{equation}	  
where $\mathbf{E_{inc,j}}$ represents the incident electric field at dipole $j$, and $\mathbf{P_{k}}$ is the polarisation of each of the other dipoles $k$ (not to be confused with the polarisation vector of the incident electric field in Eq.~\ref{polarisation_state_of_EM_field}). Note also that $k$ is defined as the wavenumber of the radiation in all equations in this paper, but also represents the an index of a dipole (e.g. dipole $k$) when used as subscript in Eq.~\ref{compact_linear}-\ref{A_jk_equation}; this is to remain consistent with the original notation of \cite{draine1994discrete}. $\mathbf{A_{jk}}$ is a tensor that calculates the contribution of the scattered electric field from all other dipoles $k$ at dipole $j$.

\begin{multline} \label{A_jk_equation}
    \mathbf{A_{jk}}=\exp\left(\frac{ikr}{r^{3}}\right)\left[k^{2} \begin{pmatrix} xx-r^{2} & xy & xy \\ yx & yy-r^{2} & yz \\ zx & zy & zz-r^{2} \end{pmatrix} \right.\\ \left.+ \frac{ikr-1}{r^{2}}\begin{pmatrix} 3xx-r^{2} & xy & xy \\ yx & 3yy-r^{2} & yz \\ zx & zy & 3zz-r^{2} \end{pmatrix}\right],
\end{multline}
where $x$, $y$ and $z$ are the relative positions of the dipoles in Cartesian coordinates, and $r=\sqrt{x^{2}+y^{2}+z^{2}}$. 

Each element of the tensor $\mathbf{A_{jk}}$ is a $3\times 3$ matrix. We further define $\mathbf{A_{jj}}=\alpha^{-1}$, which was formulated in the initial DDA theory as:

\begin{equation}
    \alpha_\mathrm{cm}=\frac{3d^{3}}{4\pi}\frac{\epsilon-1}{\epsilon+2},
\end{equation}
where $\epsilon$ is the dielectric function of the aerosol (given by $\epsilon=m^{2}/\mu$ where $\mu$ is the relative permeability of the material), and $d$ is the spacing between dipoles. This has been through several iterations of improvements to better represent discretised geometries where $d$ is not infinitely small (see Ch.3 of \citet{Yurkin_Hoekstra_2007} for a description of the historical improvements). The prescription used here is the corrected lattice dispersion relation (CLDR):

\begin{equation} \label{alpha_cldr}
    \alpha_\mathrm{CLDR}^{\mu\nu}=\frac{\alpha_\mathrm{cm}\delta_{\mu\nu}}{1+\frac{\alpha_\mathrm{cm}}{d^{3}}\left[(b_{1}+m^{2}b_{2}+m^{2}b_{3}a_{\mu}^{2})(kd)^2-\frac{2}{3}i(kd)^{3}\right]},
\end{equation}
where $b_{1}=-1.891531$, $b_{2}=0.1648469$, $b_{3}=-1.7700004$, $\mu$ and $\nu$ are indices of the $3\times3$ matrix, $\delta_{\mu\nu}$ is the delta function, $a_{1}=\cos^{2}\theta$, $a_{2}=\sin^{2}\theta \cos^{2}\phi$ and $a_{3}=\sin^{2}\theta \sin^{2}\phi$ (the angles of the incident radiation in spherical polar coordinates). For clarity, the $\mathbf{A_{jk}}$ tensor has been constructed for a very small and simplified system of 3 dipoles in Appendix \ref{appendix:expanded_DDA_matrix}.

For a given system of dipoles, we can use Eqs.~\ref{Eq:E_inc} to \ref{alpha_cldr} to calculate $\mathbf{A_{jk}}$ and $\mathbf{E_{inc,j}}$, and so the problem reduces to solving Eq. \ref{compact_linear} for a large series of $3N$ linear equations to find the polarisation states $\mathbf{P_{k}}$ (see Appendix \ref{appendix:expanded_DDA_matrix} for an expanded-form explanation of this procedure for clarity). The cross-sections are then determined using:

\begin{equation} \label{DDA_C_Ext}
C_\mathrm{ext}=\frac{4 \pi k}{\vert E_{0} \rvert^{2}} \sum_{j=1}^{N}\mathrm{Im}(E_{\mathrm{inc},j}^{*}.P_{j}),
\end{equation}

\begin{equation} \label{DDA_C_abs}
C_\mathrm{abs}=\frac{4 \pi k}{\vert E_{0} \rvert^{2}} \sum_{j=1}^{N} \{\mathrm{Im}(P_{j}.(\alpha_{j}^{-1})^{*}P_{j}^{*}) - \frac{2}{3}k^{3}\lvert P_{j} \rvert^{2}\},
\end{equation}
where in practice the electric field is normalised such that $\vert E_{0} \rvert^{2}=1$ (the initial polarisation state in Eq. \ref{polarisation_state_of_EM_field} is chosen as a unit vector). Finally, we can use Eq.~\ref{DDA_C_Ext} and \ref{DDA_C_abs} calculate $C_\mathrm{sca} = C_\mathrm{ext} - C_\mathrm{abs}$. Because the cross-sections obtained are dependent on the exact polarisation state and incident angle of radiation, the final cross-sections used in this paper are averages of two orthogonal polarisation states, and also averaged over 252 angles equally distrbibuted in 3-dimensions (see Appendix \ref{appendix:angle_averages} and \ref{appendix:polarisation_averages}for details on numerical tests that allowed us to arrive at this value, as well as the positions of angles used for averaging). In DDA, the asymmetry parameter is calculated using:
\begin{equation} \label{g_DDA}
g = \frac{1}{C_\mathrm{sca}} \int_{0}^{4\pi} \frac{\mathrm{d} C_\mathrm{sca}(\mathbf{\hat{n}},\mathbf{\hat{k}})}{\mathrm{d}\Omega}\mathbf{\hat{n}} \mathrm{d}\Omega,
\end{equation}
where $\mathbf{\hat{k}}$ is the direction of radiation propagation and $\mathrm{d}\Omega$ is the element of solid angle in scattering direction $\mathbf{\hat{n}}$. For more details on this calculation, we direct readers to the \texttt{DDSCAT} user guide \citep{draine1994discrete}.

\subsubsection{Validity} \label{DDA_validity}

Much work has been done to determine the regime in which DDA is valid, and it is generally accepted that the extinction and scattering cross-sections will be valid to within a few percent as long as the following criteria are satisfied:
\begin{enumerate}
    \item the dipole size is small compared to the wavelength:
    \begin{equation} \label{eq:validity_criteria_i}
        |m|kd<\beta, 
    \end{equation}
    where $\beta$ is a numerical factor $\approx 1$.
    \item $d$ must be small enough to describe the geometry satisfactorily.
\end{enumerate}

The numerical value of $\beta$ in Eq.~\ref{eq:validity_criteria_i} is typically set as $\approx 1$ because this ensures very high accuracy (within a few \%) for pseudospheres (which are not well-represented by cubic dipoles unless the dipole number is high) over a very wide range of wavelengths \citep{draine1994discrete,Zubko_Petrov_Grynko_Shkuratov_Okamoto_Muinonen_Nousiainen_Kimura_Yamamoto_Videen_2010}. In practical terms, condition (i) can be interpreted as an assumption that the electric field stays roughly constant within each dipole; however, the exact point at which this stops being true is not well-known. It is proposed here that condition (i) could be relaxed and DDA would still achieve reasonable accuracy over a specific set of wavelengths that we are interested in for astrophysical observation purposes; this idea is studied further in section \ref{sec:results}. 

Condition (ii) is also not well-defined; in this research, we upscale and downscale the resolution of shapes made from collections of dipoles to investigate the sensitivity of the cross-sections obtained to point (ii) above. Because the computation time strongly depends on dipole number $N$, we wish to minimise $N$ wherever possible, and a key research question becomes: for fractal aggregates, over the range of parameters that are interesting to astrophysical atmospheres, what is the minimum number of dipoles from which DDA can accurately obtain optical cross-sections, and how does that value compare to Mie and MMF theories?

\section{Methods} \label{sec:methods}

\subsection{Realistic Fractal Geometries}

\begin{figure}
    \includegraphics[width=\columnwidth]{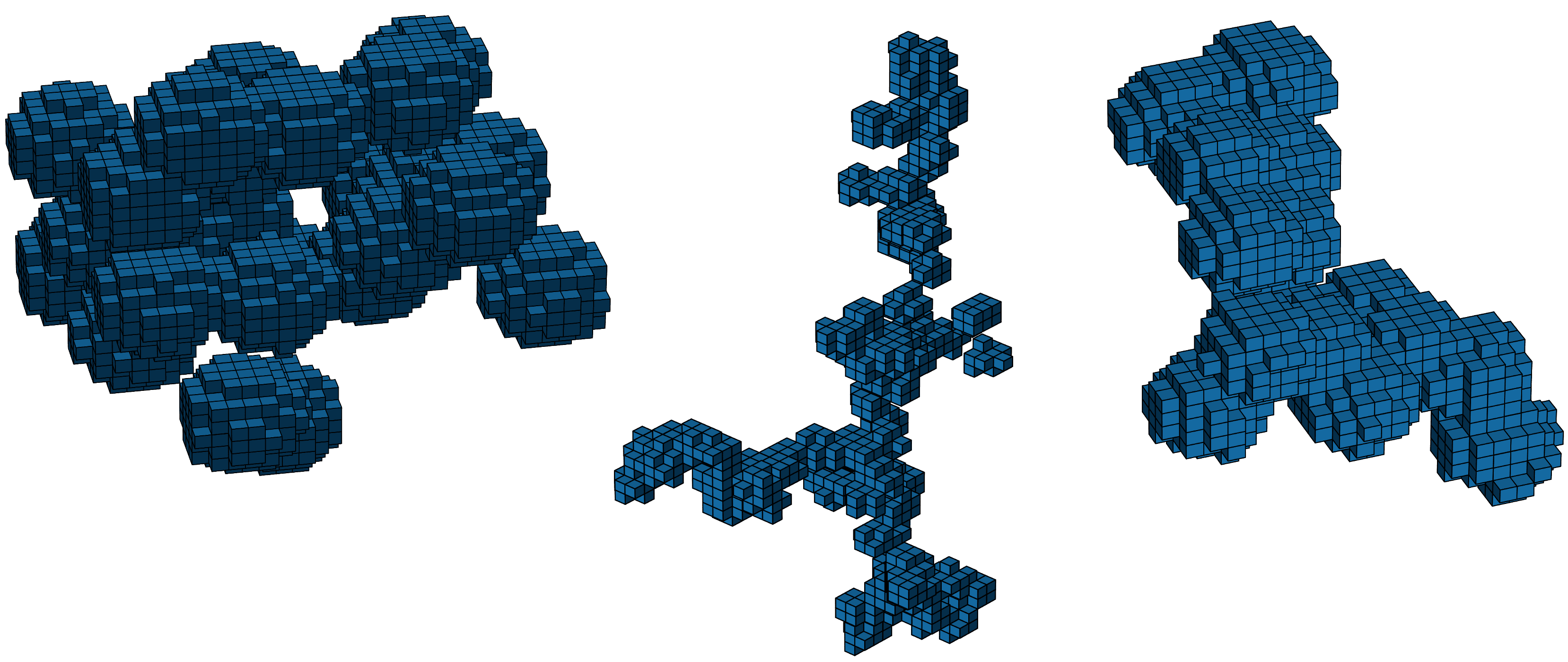}
    \caption{A selection of three of the original aerosol shape files, scanned from 3D TEM images by \citet{adachi2010shapes} and viewed in \texttt{STAG}. These specific shapes were chosen to be used in this study to represent three main different structure types: a compact 21-monomer cluster (left), a linear branched 76-monomer fractal aggregate (middle), and an elongated 15-monomer cluster (right).}
    \label{fig:realistic_fractals}
\end{figure}

In order to represent realistic aerosol geometries as accurately as possible, we use 3D models of real soot particles that were captured above Mexico city by \citet{adachi2010shapes}. These particles were scanned in a series of 2D TEM images obtained systematically along different viewing directions to create a full 3D model of their structure, for 47 different shape types. The soot structures were often combined with other materials, such as organics, and the components are separable within the files, allowing full exploration of the optical properties of multi-composite materials. For this initial study, we have restricted our analysis to the soot components only; a selection of these shape types are displayed in Figure~\ref{fig:realistic_fractals}. We chose three to analyse: one that shows particularly clear branching features, one that is more compact, and a cluster that is in-between. Novel open-source code has been developed in C and Python to visualise these shapes in 3D (\texttt{STAG}: Simulated Three-dimensional Aerosol Geometries)\footnote{\url{https://github.com/mglodge/STAG.git}}, demonstrated in Figure~\ref{fig:realistic_fractals}. Interactive versions are also available in the supplementary material and on the author's website\footnote{\url{https://www.star.bris.ac.uk/matt_lodge/home.html}}.

\subsection{\texttt{SPHERIFY}: Enabling resolution upscaling/downscaling for DDA}

As described in section \ref{DDA_validity}, condition (ii) of the validity criteria for DDA is that the shape geometry is well-represented. However, we noted that the edges of the monomers were too ``square'' compared to the curved edges of the real monomers seen in Figure~\ref{fig:TEM_fractals}, particularly for the branched fractal. Therefore, a new code (\texttt{SPHERIFY}\footnote{\url{https://github.com/mglodge/SPHERIFY.git}}) was developed to upscale the resolution. Perfect resolution upscaling is challenging, however we can quickly achieve rounder edges using quite a simple technique: double the grid size, and then interpolate between the gaps, using the two following rules:
\begin{itemize}
    \item Rule 1) Round the edges of a cube: tri-axis corners are removed and replaced with a smaller dipole at higher resolution.
    \item Rule 2) Fill the gaps of a cross: for any internal corners (as in the shape of a cross), add a smaller dipole at higher resolution.
\end{itemize}

\begin{figure*}
    \includegraphics[width=\textwidth]{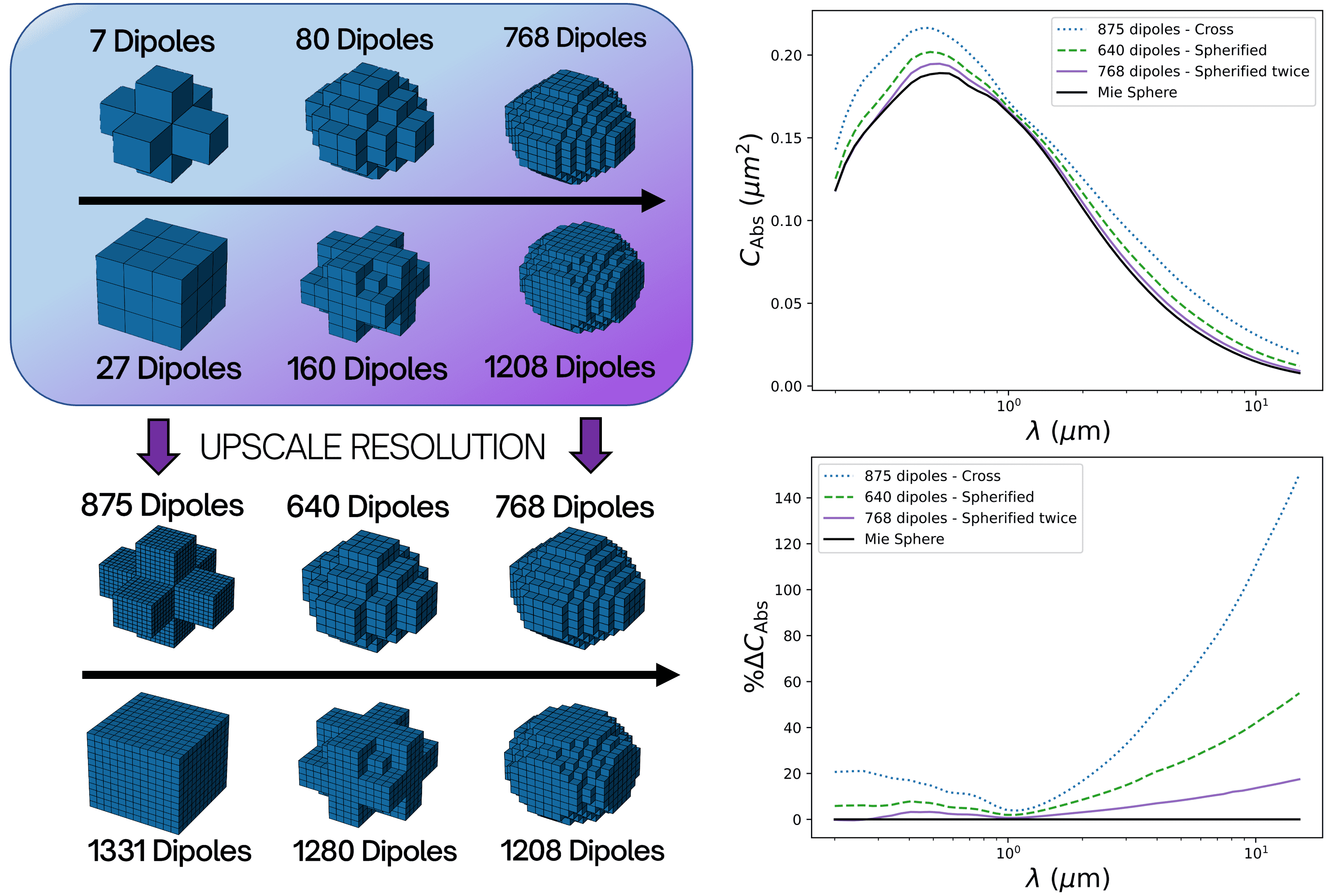}
    \caption{\textit{Top-left}: The \texttt{SPHERIFY} algorithm doubles the number of dipoles in each of the $x$, $y$, and $z$ dimensions, and then interpolates the space in between "gaps" (top), and ``rounds" off edges (bottom). The result for cube or cross shapes is a more rounded edges, at higher resolution. 
    \textit{Bottom-left:} We then increase the resolution of the shapes so that they all contain roughly the same numbers of dipoles, to allow fair comparison of assessing the shape change (independently) on the particle's scattering properties. 
    \textit{Top-right:} The absorption cross-section for the three iterations of the cross shape on the top row of the bottom-left panel. As the edges becomes more rounded; with each iteration of \texttt{SPHERIFY}, the optical profile becomes closer to that of a Mie sphere of the same volume.
    \textit{Bottom-right:} Residuals for the same graph (\% difference versus a Mie sphere), highlighting just how much closer the profile matches that of a sphere (especially at longer wavelengths).}
    \label{fig:spherify_proof}
\end{figure*}

Figure~\ref{fig:spherify_proof} (top-left) demonstrates the effects of each of these rules on two shapes. For these initial tests and proof-of-concept only, we then upscaled the number of dipoles/resolution of each shape (see bottom-left panel) before comparing their optical properties, so that the sizes of the dipoles are roughly similar, and therefore any comparison of their optical properties is more valid. The panels on the right demonstrate how the optical cross-section profiles of the cross become more similar to that of a Mie sphere (of the same volume) as the shapes become more round. The full set of data for $C_\mathrm{ext}$, $C_\mathrm{sca}$ and $C_\mathrm{abs}$ for both the cube and cross are provided in Appendix \ref{appendix:spherify_results} as more rigorous proof. It should be noted that the actual shape eventually produced by this algorithm is not a sphere, but a rhombicuboctahedron, however these graphs clearly show that one or two iterations on characteristic cubic-style DDA shape data files does produce a much better approximation of a sphere.

\begin{figure}
	\includegraphics[width=\columnwidth]{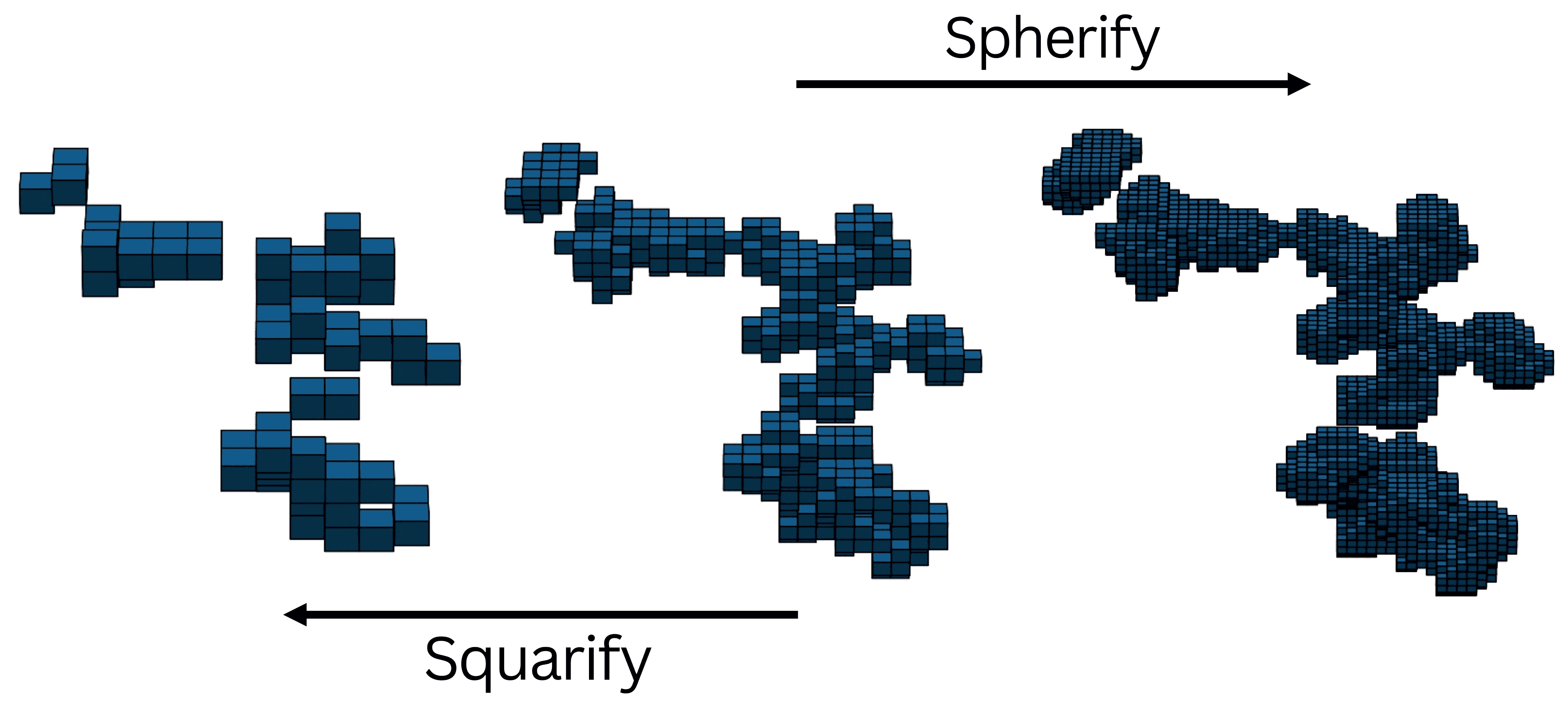}
    \caption{\textit{Center:} An original-resolution aerosol shape (536 dipoles): a small section of a real fractal soot aerosol from \citet{adachi2010shapes}, visualised in STAG. 
    \textit{Left:} The same aerosol, after one iteration of \texttt{SQUARIFY}; the number of dipoles is reduced (64 dipoles), resulting in a much lower-resolution of the same structure. 
    \textit{Right:} The same aerosol after one iteration through \texttt{SPHERIFY}; the resolution is doubled (4597 dipoles), and the edges are now much more rounded, better representing the curved edges of the real particles.}
    \label{fig:spherify_fractal}
\end{figure}

An additional consequence of spherification is that the dipole number $N$ increases (roughly by a factor of $2^{3}=8$, because we double the resolution in each of the x, y and z dimensions). This significantly increases the computation time for DDA, for which the matrix size is directly proportional to $N^{2}$, therefore one iteration of \texttt{SPHERIFY} increases the computation time of a simple implementation of DDA by a factor of $\approx64$. The effect can be mitigated by using more complex DDA implementations, for example utilising fast Fourier transforms to make computation time dependent on $O(N\log N)$ rather than $O(N^{2}$) \citep{draine1994discrete}, but the general rule is still true: smaller numbers of $N$ are always faster to calculate. For this reason, it is also interesting to decrease the resolution of the shapes, to ask: ``What is the smallest number of dipoles $N$ that will still give accurate cross-sections?'. This is achieved through ``squarification'': each 2x2x2 cube of dipoles in the original shape file becomes a single dipole in low-resolution space. From the original eight dipoles, if at least four out of eight dipoles are occupied, then it remains a dipole in the new low-resolution space. Otherwise, it becomes empty. The effects of these two codes on a small section of a realistic fractal shape is shown in Figure~\ref{fig:spherify_fractal}.

\subsection{Comparisons Between Models} \label{section:comparisons_between_models}

To compare between Mie, MMF and DDA models, we conserve volume and mass between them; that is, we assume that there is a fixed spherical mass of aerosol material (a Mie sphere), and we explore how its optical properties change when we model the same mass distributed instead into one of the fractal shapes, such as those in Figure~\ref{fig:realistic_fractals}. It is assumed that the density of the material does not change. The total volume of the shape in a perfectly spherical Mie sphere is given by
\begin{equation} \label{eq:mie_volume}
    V=\frac{4}{3}\pi R_\mathrm{mie}^3,
\end{equation}
whereas in DDA it is equal to the number of dipoles ($N$) multiplied by the volume of each one: $V=Nd^{3}$, where $d$ is the spacing between dipoles. This means that for a particular particle size distribution (from which we choose a value of $R_\mathrm{mie}$ to analyse), we can calculate $d$ by equating these two equations for volume and rearranging:
\begin{equation} \label{eq:DDA_dipole_size_d}
    d = \sqrt[3]{\frac{4 \pi}{3N}} R_\mathrm{mie}.
\end{equation}

When comparing these two models to MMF, Eq.~\ref{fractal_equation} initially appears to contain two free parameters that describe the shape (the fractal prefactor $k_{0}$ and fractal dimension $d_{f}$); however Eq. 2 in \citet{Tazaki_2021} links these together with an approximation (that is correct to within 5\% for all of the structure-types explored here):
\begin{equation} 
    k_{0} \approx 0.716(1-d_{f}) + \sqrt{3}.
\end{equation}
Substituting this into Eq.~\ref{fractal_equation} gives:
\begin{equation} \label{new_fractal_equation}
    N_\mathrm{mon} \approx \left[2.448 - 0.716d_{f}\right] \left(\frac{R_{g}}{R_{0}} \right)^{d_{f}}.
\end{equation}

The number of monomers ($N_\mathrm{mon}$) can simply be counted visually for a particular shape type (for example, see Figure~\ref{fig:realistic_fractals}). The radius of gyration ($R_{g}$) is defined as the radial distance to a point which would have a moment of inertia the same as the body's actual distribution of mass, if the total mass of the body were concentrated there. Therefore, using a shape that is composed of a high number of dipoles, this can be computed from $N$ dipoles of mass $m_{i}$ at radius $r_{i}$ from the center of mass using:
\begin{equation} \label{Eq:radius_of_gyration}
    R_{g} = \sqrt{\frac{m_{1}r_{1}^{2} + m_{2}r_{2}^{2} + ... + m_{N}r_{N}^{2}}{\sum\limits_{i=1}^{N} m_{i}}}.
\end{equation}

Finally, the radius of the monomers ($R_{0}$) can be determined through conservation of volume, and utilising the concept that all monomers in MMF are the same size. In MMF, the total volume of $N_\mathrm{mon}$ monomers of radius $R_{0}$ is $V=N_\mathrm{mon} \frac{4}{3}\pi R_{0}^{3}$. Equating this to the equation for volume of a Mie sphere (Equation~\ref{eq:mie_volume}) gives
\begin{equation} \label{Eq:radius_of_monomer}
    R_{0}=\frac{R_\mathrm{mie}}{\sqrt[3]{N_\mathrm{mon}}}.
\end{equation}

Substituting values from Eq.~\ref{Eq:radius_of_gyration} and \ref{Eq:radius_of_monomer} into Eq.~\ref{new_fractal_equation} allows us to solve for $d_{f}$. The analytical solution is complex, and the derivation is lengthy, however in practise it is also very easy and fast to solve numerically to good accuracy since we know that the value of $d_{f}$ should lie somewhere between $1-3$. With this value obtained, all equations for MMF can now be computed and compared to those of Mie and DDA. The exact values obtained using this methodology for the three fractal shape types used in this study are listed in Table~\ref{table:fractal_parameters}. As an additional note, having the actual shape files allowed direct calculation of structure factor. This only resulted in very minor differences in the cross-sections calculated using MMF, and thus the main results of the paper were unchanged.

\begin{table}
	\centering
	\caption{Table showing the MMF parameters used for the number of monomers $N$, fractal dimension $d_{f}$, and fractal prefactor $k_0$ for all three shapes used in this study. 3D models of each one are shown in Figure~\ref{fig:realistic_fractals}. The radius parameters depend on the actual size of the particles analysed, and can be calculated using Eq.~\ref{new_fractal_equation}-\ref{Eq:radius_of_monomer}.}
	\label{table:fractal_parameters}
	\begin{tabular}{cccc} % four columns, alignment for each
		\hline
		Shape type & $N$ & $d_{f}$ & $k_{0}$\\
		\hline
		CC (compact cluster) & 21 & 2.529 & 1.224\\
		LB (linear branched) & 76 & 1.811 & 1.151\\
		EC (elongated cluster) & 15 & 2.537 & 0.845\\
		\hline
	\end{tabular}
\end{table}

The wavelength-dependent refractive index profile used in all three of the models is the same. Three different profiles were explored for use in this study: one for soot \citep{chang1990determination} similar to those used by \citet{gao2020aerosol,morley2013quantitatively}, the original tholins dataset by \cite{khare1984optical} (used in \citealt{wolf2010fractal}), and an update of this profile using lab-created photochemical hazes that are a good representation of what we might expect to find in super-Earth and mini-Neptune atmospheres \citep{he2018laboratory}. For all graphs, figures and tables in the main body of this paper, we use the soot profile, because the consequences of using such a variable profile on fractal aerosols are less-frequently studied. The one exception is in Appendix \ref{appendix:wolf_and_toon_update}, where we use \citet{khare1984optical} to recreate the original \citet{wolf2010fractal} data, but extend out to longer wavelengths than in the original paper. In each case, it is important to note the refractive index profile is the single input to the models that differentiates between different chemical compositions, and in reality any of these profiles (and a wide range of others) could be found in exoplanet atmospheres. The variation in observable outputs as a result of these different modelling assumptions is of great interest to the exoplanet atmosphere community.

\begin{figure}
    \includegraphics[width=\columnwidth]{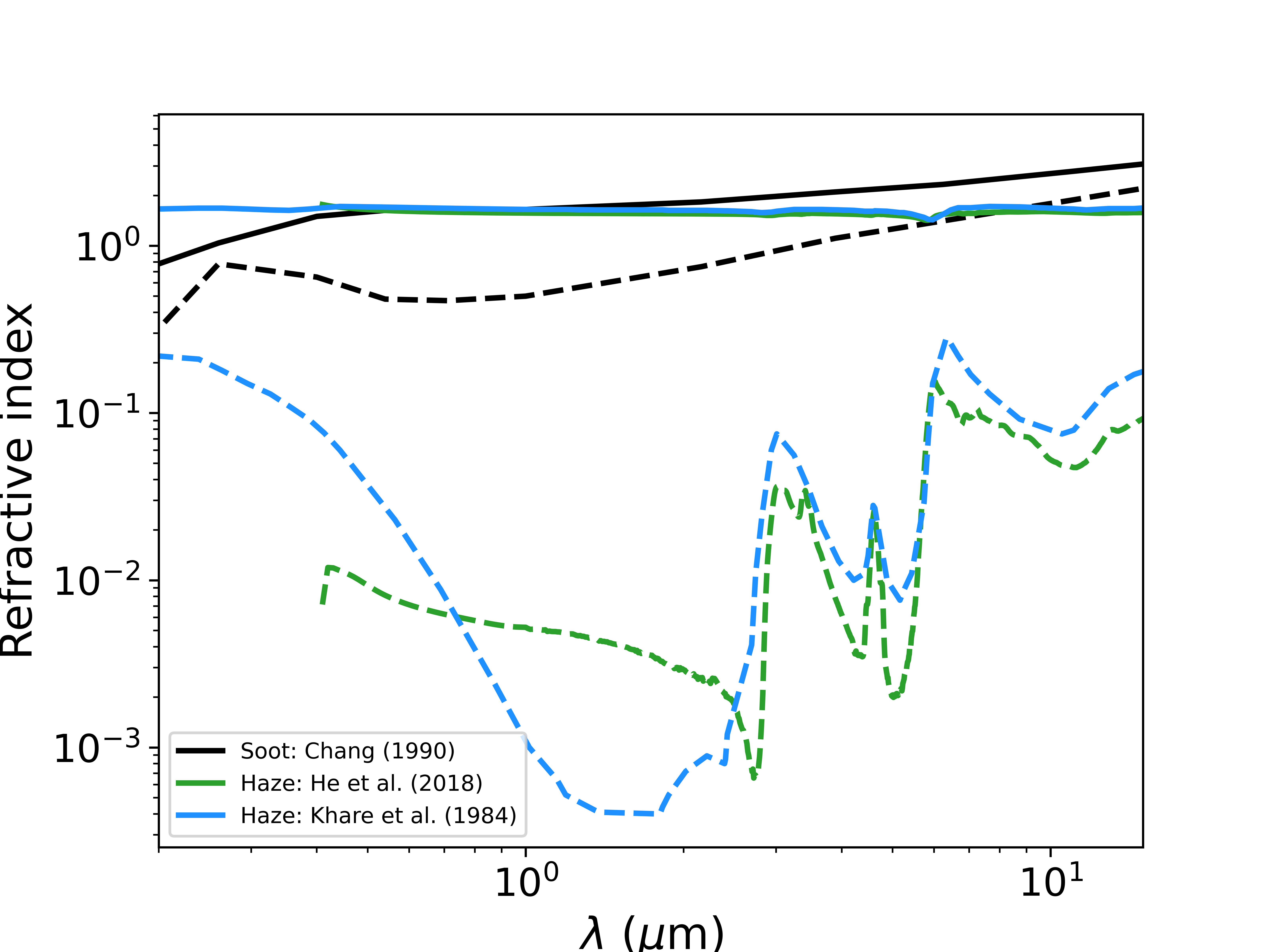}
    \caption{Empirically obtained real ($n$: solid) and imaginary ($k$: dashed) components of the refractive index profiles for soots and hazes used in this study. The haze data \citep{he2018laboratory} considered wavelengths above 0.4 $\mu$m only (green solid and dashed lines). The soot profile (black) was used in all results, figures and tables from this point onwards.}
    \label{fig:refractive_index_profiles}
\end{figure}

\subsection{Computational Methods} \label{sec:computational_methods}

Optimised codes exist and are freely available for calculating Mie, MMF and DDA cross-sections (\texttt{BHMIE} \citep{bohren2008absorption}, \texttt{optool} \citep{2021ascl.soft04010D} and \texttt{DDSCAT} \citep{draine1994discrete} respectively); however because we wished to compare these models in great detail and a wide array of specific and consistent input parameters, and then integrate these calculations as part of a wider radiative transfer package, we wrote new C code \texttt{CORAL} (Comparison Of Radiative AnaLysis)\footnote{\url{https://github.com/mglodge/CORAL.git}} to compute all three models. Although not quite as fast as the individually optimised codes above, it has been thoroughly tested and benchmarked (see Appendix~\ref{appendix:benchmark_tests}) for a wide range of parameters and agrees to at least 4 significant figures for all three scattering parameters in all cases tested (any difference beyond this is expected, due to very slightly different implementations of numerical methods, for example numerical integration and linear equation solvers). All codes used have been made freely available, and further testing by others is freely encouraged.

\section{Results} \label{sec:results}

It is important to note, with justification, which we believe to be the most accurate model out of Mie, MMF and DDA. It should be highlighted that it is very difficult, if not impossible, to measure the optical properties of a specific complex-shaped submicron-sized aerosol particle in a laboratory experiment to verify the results directly. In this study, we follow aerosol literature in trusting that the most accurate values for cross-sections are obtained using DDA, and specifically the model with the highest number of dipoles (because as discussed in section \ref{Theory:DDA}, the accuracy of theory improves as $N\rightarrow\infty$); the highest $N$ for each of the three shape types is our benchmark for accuracy.

\subsection{Wolf and Toon (2010) - revisited with a soot profile} \label{results:wolf_and_toon}

\begin{figure}
    \includegraphics[width=\columnwidth]{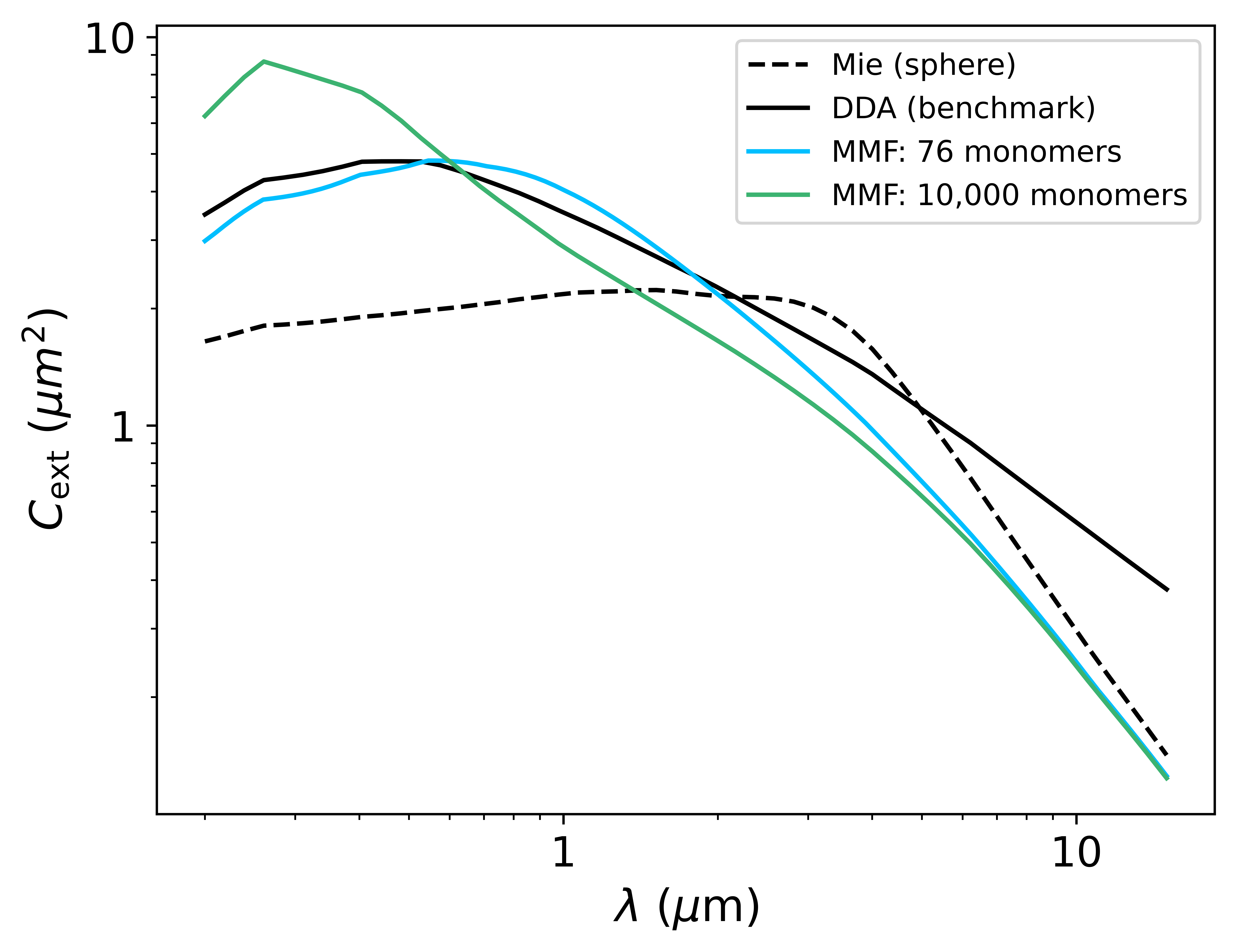}
    \caption{Extinction cross-sections are plotted for a $0.5 \mu \mathrm{m}$ branched fractal (center, Figure \ref{fig:realistic_fractals}) for Mie, MMF and DDA, using the soot refractive index profile from \citet{chang1990determination}. The shape was assumed to have 76 monomers by visual inspection. An alternative profile for MMF, ``forcing'' a higher value of 10,000 monomers, is also plotted (in green). DDA demonstrates that there is more extinction than one would predict when assuming the particle is spherical, at almost all wavelengths (except between $2-4 \upmu m$).}
    \label{fig:DDA_v_MMF_v_Mie_0.5um_soot}
\end{figure}

Figure \ref{fig:DDA_v_MMF_v_Mie_0.5um_soot} demonstrates the extinction cross-section for a linear branched hydrocarbon, using the soot profile from \citet{chang1990determination}. Here we determined that the fractal aggregate was composed of 76 monomers by visual inspection, and using the methodology outlined in section \ref{section:comparisons_between_models}, we obtained the MMF parameters in Table \ref{table:fractal_parameters}. Comparing just MMF (76 monomers) and Mie theories, we reach the same conclusion as \cite{wolf2010fractal}; there is more extinction for fractal shapes than for spheres at the shortest wavelengths, and less extinction at longer wavelengths. However, the differences are much smaller than in \citet{wolf2010fractal}. The largest differences were seen for high monomer numbers; we therefore artificially increased this value by assuming that our fractal is made of a higher number of smaller monomers (i.e. made of 10,000 monomers of 23~nm radius, instead of 76 monomers of 118~nm radius; both of which would have identical volume). Interestingly, this produces results closer to the larger extinction values seen in \citet{wolf2010fractal}, though it should be noted that the methodologies used to increase monomer number in the two studies are not the same. In \cite{wolf2010fractal}, the number of monomers was increased by adding extra monomers of the same radius and thus the total particle size/mass also increased. Here, we assumed that the particle size/mass was constant, but simply chose it to be made of a much higher number of smaller particles, and note that the optical properties can change by orders of magnitude. When comparing DDA and MMF using conservation of volume for these shape files, the number of monomers is judged by eye, which means that values can vary, unless one obtains prior chemistry/particle formation information. This section highlights that when using this method of comparison, the results of MMF can be sensitive to the chosen value for number of monomers, and thus this value should be considered with care. The first case (76 monomers) is expected to be the most accurate here, and fits closer to our benchmark model for accuracy, which is the DDA curve.

Analysing the short wavelengths of the DDA profile in more detail (the portion studied in \citet{wolf2010fractal}), the $0.2-4~\upmu$m region gives identical conclusions to MMF (it shows that there is more extinction than for a Mie sphere at shorter wavelengths, and less extinction at longer wavelengths). However, the interesting result is that extending beyond this wavelength range reveals that the fractal aggregates quickly become more absorptive than spheres again (when $\lambda>4.5~\upmu\mathrm{m}$). It was noticed that the refractive index becomes relatively large ($m\approx3 + 2i$) at the highest wavelengths studied (15~$\upmu$m), and DDA has been known to cause errors for these cases \citep{draine1994discrete}. However, to quantify this error, a 5616-dipole pseudosphere was analysed (which should have the same optical properties as a Mie sphere) with the same refractive index profile. At 15~$\upmu$m, the extinction cross-section of the pseudosphere was only 8.6\% larger using DDA than when compared to Mie, whereas for the linear branched fractal the cross-section is 267\% larger than Mie at 15~$\upmu$m (see Figure \ref{fig:DDA_v_MMF_v_Mie_0.5um_soot}). We are therefore confident that any error introduced by higher refractive indices is small, and cannot explain the large difference in cross-section alone.

This same effect (fractal shapes showing highly increased absorption at long wavelengths) also happens even when we used the original flatter haze profile ($m\approx1.6+0.1i$) studied in \citet{wolf2010fractal}. However, if a flatter profile is used, the cross-sections are much smaller and almost negligible in this region, and so this may be why the effect has not been noticed before (see Appendix \ref{appendix:wolf_and_toon_update}).

The conclusion is that soot-based fractal aggregates absorb much more than their spherical counterparts at almost all wavelengths; this is a valuable addition to the \citet{wolf2010fractal} result that particles with flatter haze profiles may be more transparent to long-wavelength radiation, and we demonstrate that the opposite can also be true. We highlight that there is not a one-size-fits-all rule; the absorptive and transmittal nature of the fractal aggregate seems to depend on the exact composition of the material and thus the refractive index profile used. MMF does not predict higher absorption at longer wavelengths, however it is known to underestimate cross-sections in this region \citep{Tazaki_Tanaka_2018}. Therefore, it is important that a rigorous analysis such as DDA is used to probe these wavelengths.

\subsection{Comparing Mie, MMF and DDA with a variety of aerosol resolutions/dipole numbers}

\begin{figure}
    \includegraphics[width=\columnwidth]{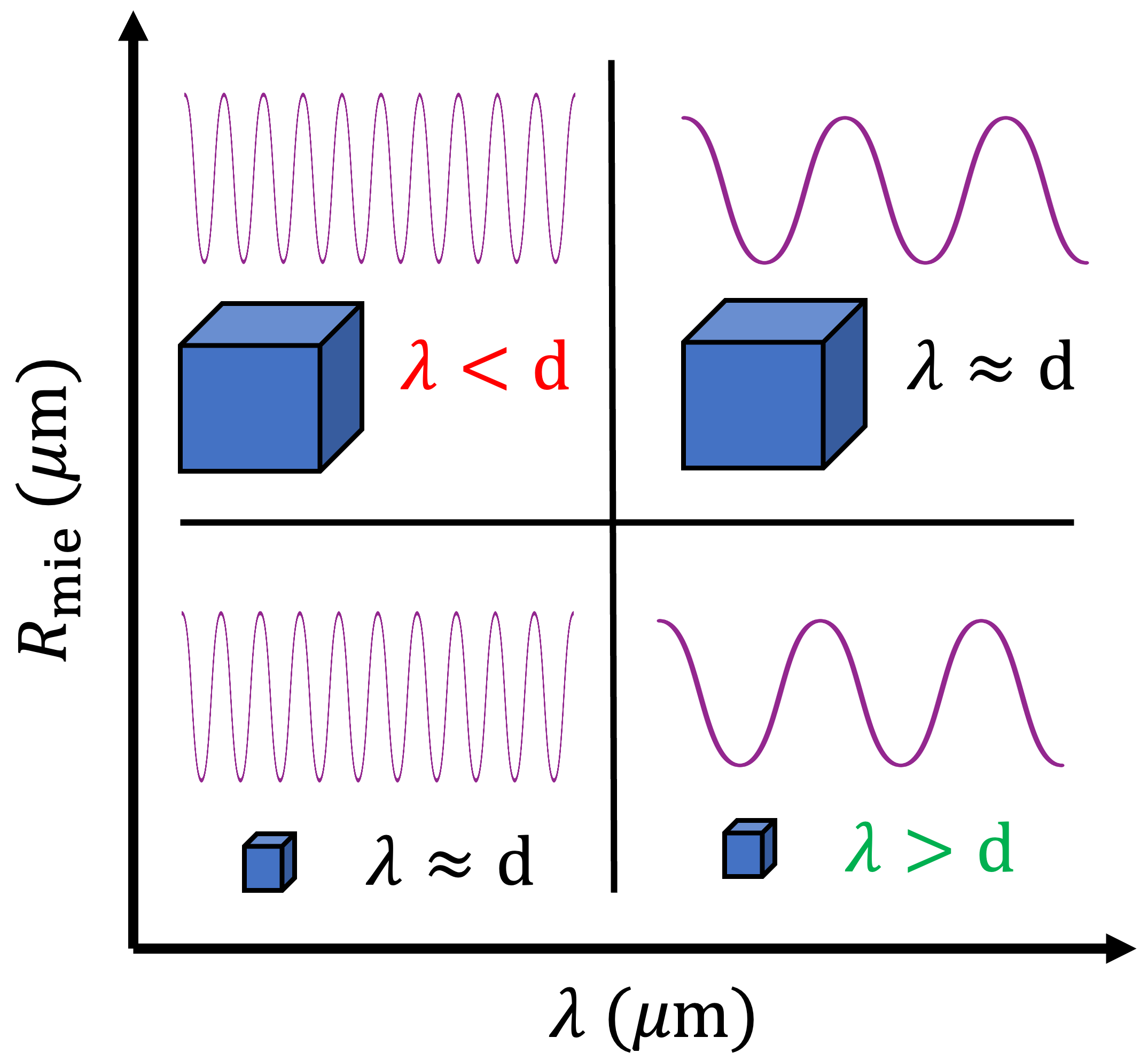}
    \caption{For DDA to be valid, the wavelength must be larger than the dipole size because the electric field is assumed to be constant within each dipole. Bottom-right: the region where DDA is most valid -- a long wavelength relative to a small dipole. Top-right: if the particle is larger, the dipole is larger, and eventually the same size as the wavelength. The electric field is no longer constant within the dipole as they are roughly the same size. Top-left: If the particle is larger (and the dipoles are larger), as well as the wavelength being shorter, DDA will no longer be valid because the electric field changes greatly within the dipoles. Bottom-left: if the wavelength is short, but the dipole is also small, eventually DDA can stop being valid when the two become comparable in size.}
    \label{fig:parameter_space_explanation}
\end{figure}

Figure \ref{fig:parameter_space_explanation} shows the general requirements for DDA to be a valid approximation, and the effects are studied in detail here by modelling each aerosol at a variety of resolutions and particle sizes.

\begin{figure*}
    \includegraphics[width=\textwidth]{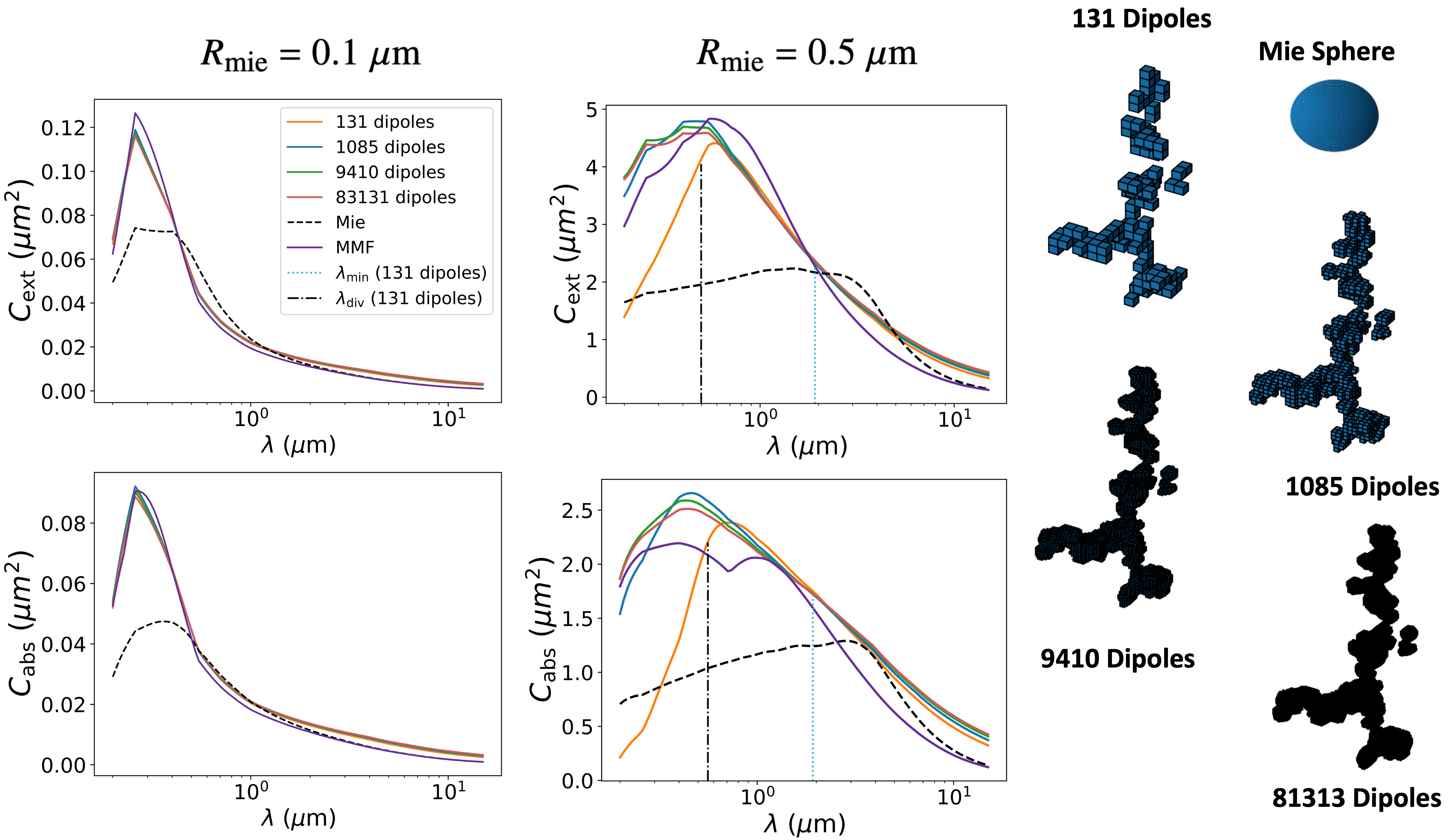}
    \caption{\textit{Left column:} Extinction and absorption cross-sections as a function of wavelength for each branched fractal structure with an equivalent Mie radius of 0.1 $\upmu$m.
    \textit{Center column:} The same graphs but with a larger equivalent Mie radius of 0.5 $\upmu$m. The wavelength below which DDA is expected to become invalid ($\lambda_\mathrm{min}$, assuming that $\beta=1$), is calculated for the 131-dipole shape and highlighted as the dotted cyan line. The \textit{actual} wavelength where DDA begins to diverge from the correct result is highlighted as the dash-dotted black line ($\lambda_\mathrm{div}$). This wavelength is calculated but not shown for the 1085-dipole shape too, tabulated in Table~\ref{table:Beta_and_lambda_min_values}. The 9410-dipole shape does not have a $\lambda_\mathrm{div}$ value, as it never diverges from the benchmark by more than 10\%.
    \textit{Right:} The original aerosol (1085 dipoles), the decreased resolution version using \texttt{SQUARIFY} (131 dipoles), and an enhanced resolution version using \texttt{SPHERIFY} once (9410 dipoles) and twice in a row (81,313 dipoles). A Mie sphere, equivalent by volume to each of the other shapes, is also shown (to scale). Residuals are shown in Fig.~\ref{fig:residuals_main}.}
    \label{fig:results_branched}
\end{figure*}

\begin{figure*}
    \includegraphics[width=\textwidth]{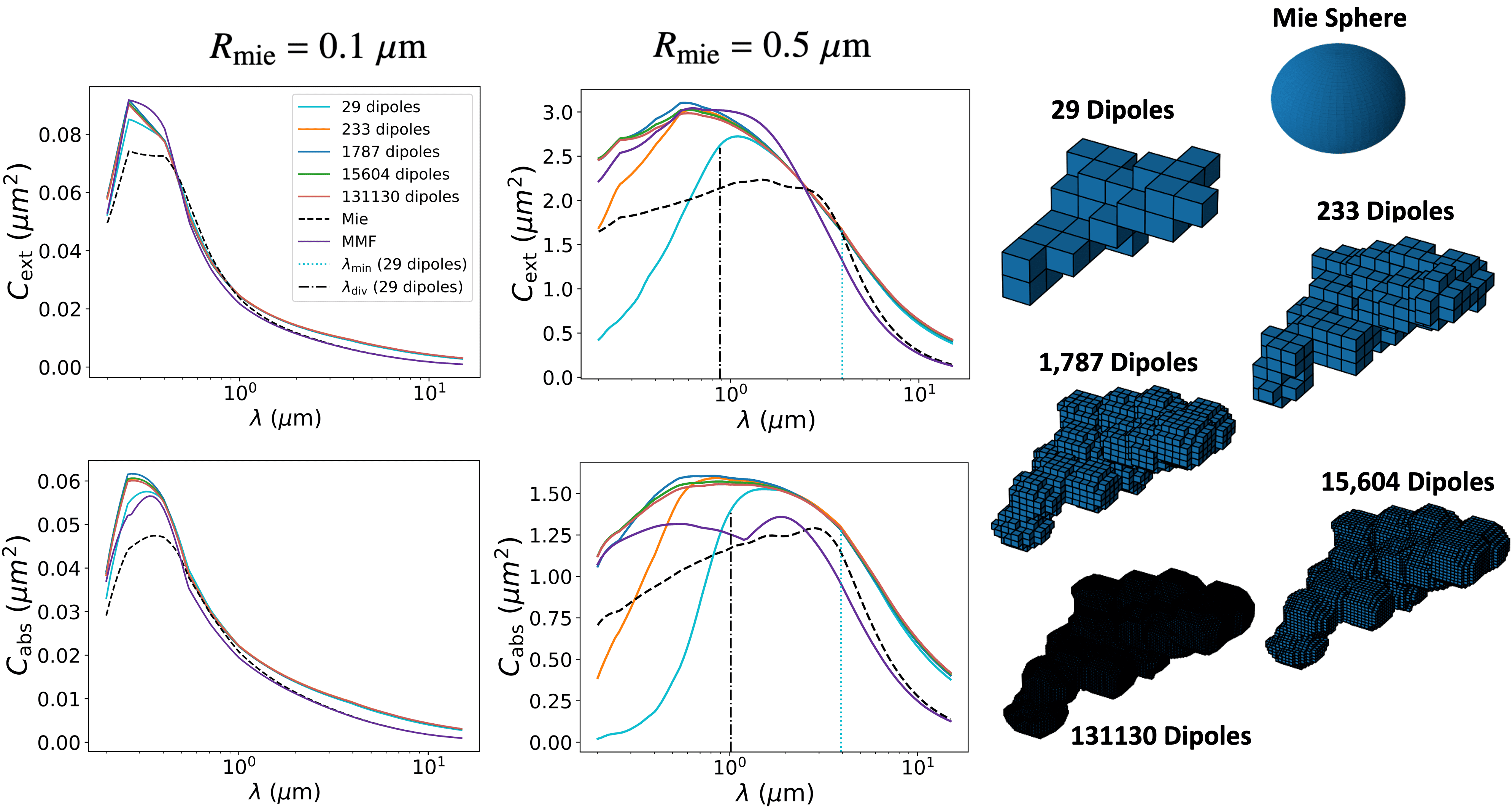}
    \caption{\textit{Left and center columns:} As in Figure~\ref{fig:results_branched} but for an elongated cluster. The wavelength below which DDA is expected to become invalid ($\lambda_\mathrm{min}$, assuming that $\beta=1$), is calculated for the 29-dipole shape and highlighted as the dotted cyan line. The \textit{actual} wavelength where DDA begins to diverge from the correct result is highlighted as the dash-dotted black line ($\lambda_\mathrm{div}$). This wavelength is calculated but not shown for the 233 and 1787-dipole shapes too, tabulated in Table~\ref{table:Beta_and_lambda_min_values}. The 15,604-dipole shape does not have a $\lambda_\mathrm{div}$ value, as it never diverges from the benchmark by more than 10\%.
    \textit{Right:} The original aerosol (1787 dipoles), the enhanced resolution version using \texttt{SPHERIFY} once (15,604 dipoles) and twice in a row (131,130 dipoles), and the decreased resolution versions using \texttt{SQUARIFY} once (233 dipoles) and twice in a row (29 dipoles). A Mie sphere, equivalent by volume to each of the other shapes, is also shown (to scale). Residuals are shown in Fig.~\ref{fig:residuals_main}.}
    \label{fig:results_elongated}
\end{figure*}

\begin{figure*}
    \includegraphics[width=\textwidth]{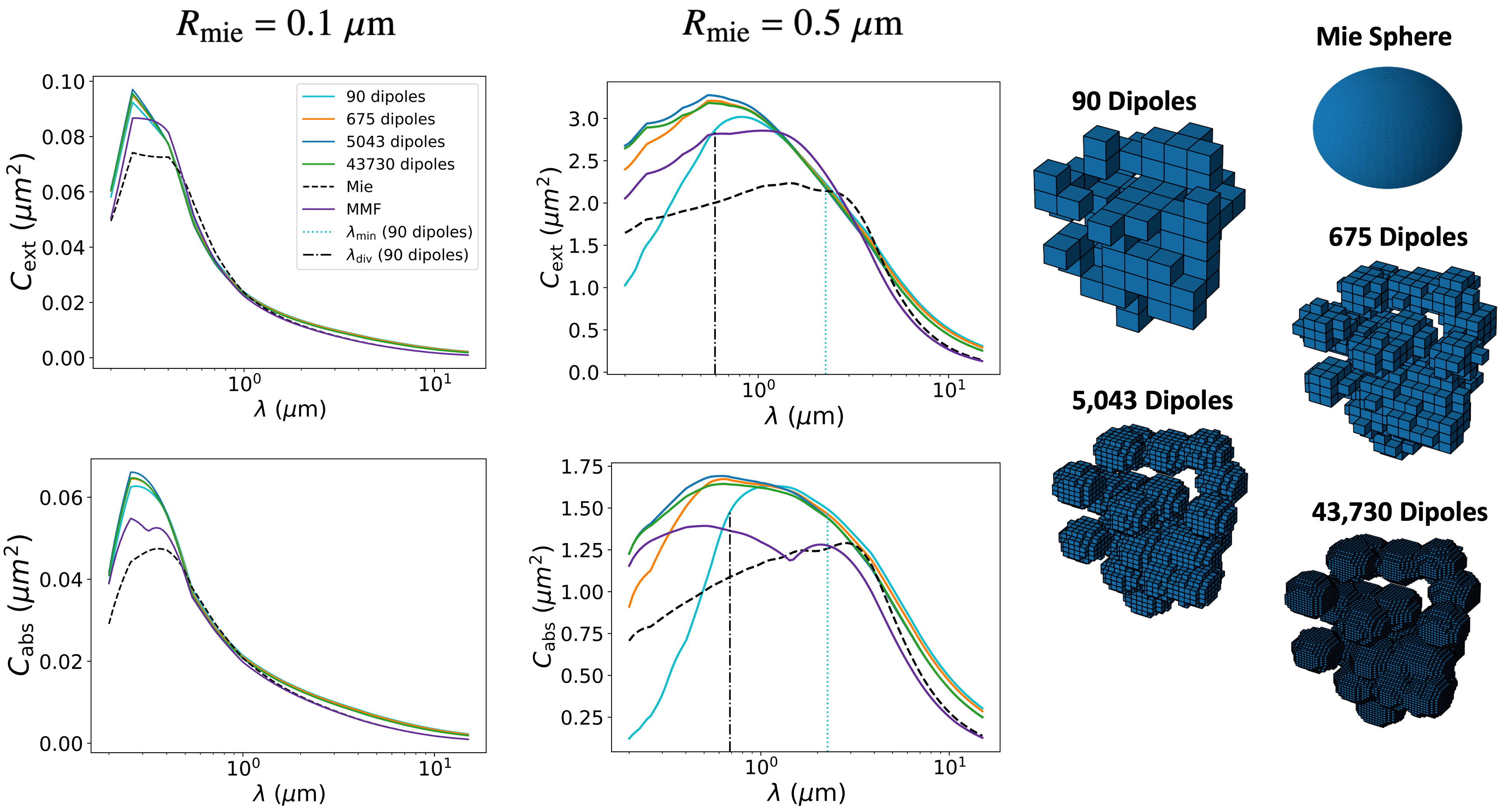}
    \caption{\textit{Left and center columns:} As in Figure~\ref{fig:results_branched} and \ref{fig:results_elongated} but for a compact cluster. The wavelength below which DDA is expected to become invalid ($\lambda_\mathrm{min}$, assuming that $\beta=1$), is calculated for the 90-dipole shape and highlighted as the dotted cyan line. The \textit{actual} wavelength where DDA begins to diverge from the correct result is highlighted as the dash-dotted black line ($\lambda_\mathrm{div}$). This wavelength is calculated but not shown for the 675-dipole shape too, tabulated in Table~\ref{table:Beta_and_lambda_min_values}. The 5043-dipole shape does not have a $\lambda_\mathrm{div}$ value, as it never diverges from the benchmark by more than 10\%. \textit{Right:} The original aerosol (5043 dipoles), the enhanced resolution version using \texttt{SPHERIFY} (43,730 dipoles), and the decreased resolution versions using \texttt{SQUARIFY} once (675 dipoles) and twice in a row (90 dipoles). A Mie sphere, equivalent by volume to each of the other shapes, is also shown (to scale). Residuals are shown in Fig.~\ref{fig:residuals_main}.}
    \label{fig:results_compact}
\end{figure*}

Figure~\ref{fig:results_branched} (left) shows the Extinction and Absorption cross-sections for aerosol particles with $R_\mathrm{mie}=0.1~\upmu$m radii; this clearly demonstrates that for this wavelength and particle radius, the cross-sections obtained using DDA are actually independent of the number of dipoles used to represent them. However, the values for the 131 dipole version were roughly $10^6$ times faster to compute than for the 81313 dipole particle. Therefore, for small particles, DDA can be used at a very low resolution and still obtain accurate results (much more quickly); there is no need to use high resolution/dipole numbers/computation time (comparisons of dipole number, computation times and \% accuracy are listed in Appendix~\ref{appendix:speed_tests}). Similar results are seen for the $0.1~\upmu$m radii particles for different shape structure types in Figure~\ref{fig:results_elongated} and Figure~\ref{fig:results_compact}, though perhaps with some increased divergence from the benchmark 131,130 and 43,730 dipole models in each case (however, even the lowest-resolution DDA was significantly more accurate than MMF). The long-wavelength limit is especially important to analyse, as it is known that MMF can significantly underestimate the absorption and extinction cross-sections at longer wavelengths by $\approx 40\%$ (see Figure 4 and Section 5 of \citet{Tazaki_Tanaka_2018} for a comprehensive analysis, including a discussion of the physical rationale for this occurring). The cross-sections obtained by DDA are in agreement with this analysis, being about twice as large as those predicted by MMF and Mie in Figure~\ref{fig:results_branched} (similar to those obtained by the T-matrix method in \citealt{Tazaki_Tanaka_2018}) -- this is more clearly seen in the residuals, provided in Appendix \ref{appendix:residuals} of this paper. \citet{Tazaki_Tanaka_2018} proposed that representing the structure more finely (in particular, looking at the contact between monomer surfaces) could explain enhanced absorption at long wavelengths. It is therefore interesting that for DDA, a more coarse (low-resolution) representation of the structure is successful in producing accurate results. We assume that this is because even for a small number of dipoles, DDA still uses accurate information about the dipole distribution/particle shape. This is an especially significant result, because if low-resolution DDA can provide accurate cross-sections at longer wavelengths (as shown here), it may be the fastest theoretical method to be able to do so, at least for structures with small monomer numbers that can described well by low numbers of dipoles. However, more research on a wider range of structure and material types is required.

For the larger particle sizes of $R_\mathrm{mie}=0.5~\upmu$m radii in Figure~\ref{fig:results_branched} (center), the effect of validity criteria (i) ($|m|kd< \beta $) in section \ref{DDA_validity} becomes clear. Beginning in the long wavelength-limit on the right of the figure, the wavelength is much larger than the dipole size. Decreasing in wavelength from this point, eventually the wavelength and dipole size are comparable, and DDA is no longer a valid approximation. This happens for the lowest resolution shapes first (131 dipoles in Figure~\ref{fig:results_branched}), because their dipole sizes are largest. This is shown visually in Figure~\ref{fig:parameter_space_explanation}.

Historically, this specific point at which DDA becomes invalid has been poorly-constrained. However, Eq.~\ref{eq:validity_criteria_i} from validity criteria (i) can be rearranged to determine the minimum wavelength where the condition is \textit{expected} to be obeyed for an assumed numerical value of $\beta$:
\begin{equation} \label{eq:validity_criteria_with_beta}
    \lambda_\mathrm{min}=\frac{2\pi|m(\lambda)|d}{\beta}.
\end{equation}
Initially, we calculate $\lambda_\mathrm{min}$ assuming the conservative numerical factor of $\beta=1$ (as in previous studies -- see section \ref{DDA_validity}). For example, the values of $\lambda_\mathrm{min}$ for the 131, 1085 and 9410 dipole models are $1.918~\upmu$m, $0.837~\upmu$m and $0.380~\upmu$m respectively (there is no $\lambda_\mathrm{min}$ for the 81,313 dipole shape, because for $\beta=1$, the validity condition is always true over the entire wavelength range). At wavelengths smaller than $\lambda_\mathrm{min}$, each respective DDA model should (in theory) no longer be valid; the value of $\lambda_\mathrm{min}$ is shown as the dashed line in Figure~\ref{fig:results_branched} (centre) for the 131 dipole shape only (for clarity). However, it is shown here that the low and medium-resolution models actually maintain accuracy at much lower wavelengths than predicted by $\lambda_\mathrm{min}$ in all cases (for all resolutions, and all shape types). For example, while $\lambda_\mathrm{min}=1.918~\upmu$m for the 131 dipole model, the extinction cross-sections actually stay accurate to within 10 \% of the benchmark 81,313 dipole model for wavelengths as low as $0.497~\upmu$m (marked as the dot-dashed line). 

Similar results are seen for all other resolutions and shape-types; just to pick one example, $\lambda_\mathrm{min}=2.261~\upmu$m for the 90-dipole compact cluster (and below this point DDA would not be expected to be valid), however the results for $C_\mathrm{ext}$, $C_\mathrm{sca}$ and $C_\mathrm{abs}$ stay accurate until the much lower values of 0.593, 0.295 and 0.685 $\upmu$m respectively. We suggest that for the specific range of parameters and shape types that we might expect to observe in exoplanet atmospheres, DDA would still be able to accurately determine accurate cross-sections below the wavelengths previously assumed to be the minimum, where $\beta=1$. From these curves, we can determine the actual wavelength at which the profile of each of the aerosols diverges from the values obtained by the benchmark (high dipole) model by a certain amount; we define this point as $\lambda_\mathrm{div}$. The exact tolerance that we allow for divergence is somewhat arbitrary, but here we allow a maximum deviation of 10\% from the benchmark; this allows for some small natural differences in the resonances obtained by difference shapes formed at different resolutions, while still ensuring reasonable accuracy at the lowest wavelengths. It should be noted that 10\% is the maximum error for the lowest wavelengths; the actual error for most of the wavelength range studied will usually be much lower than this. Once the value of $\lambda_\mathrm{div}$ is determined for each shape, it can be substituted back into a rearranged form of Eq.~\ref{eq:validity_criteria_with_beta} to numerically determine a new value of $\beta$ for each shape and at each of the studied resolutions:

\begin{equation} \label{eq:validity_criteria_solved_for_beta}
    \beta=\frac{2\pi|m(\lambda)|d}{\lambda_\mathrm{div}}.
\end{equation}
This numerically-determined $\beta$ represents a new suggested value for the validity criteria for DDA, determined numerically for each shape, allowing a tolerance of 10\% from the true values for the specific parameter space that we are interested in for exoplanet observations.

Figures~ \ref{fig:results_branched}-\ref{fig:results_compact} demonstrate that the exact same effect occurs for each shape; the lowest-resolution models each eventually reach a wavelength below which their results are no longer valid (dot-dashed line), which is always much lower than the minimum wavelength predicted when $\beta=1$ (dashed line). This is best summarised numerically in Table~\ref{table:Beta_and_lambda_min_values}. This table also demonstrates that for this parameter space, $\beta>2$ in all cases for both $C_\mathrm{ext}$ and $C_\mathrm{abs}$, and for $C_\mathrm{sca}$ the profiles barely diverge at all (even at the lowest resolutions where $\beta>4$). Also of note is the average \% error (an average taken above $\lambda_\mathrm{div}$ only) tabulated for each dipole number; although the error does tend to increase as the resolution is lowered, the average errors for all three cross-sections are much smaller than that of Mie and MMF for all three shape types. This is also visually demonstrated in Figures~\ref{fig:results_branched}-\ref{fig:results_compact}.

\begin{table*}
    \centering
	\caption{Lists the predicted wavelengths below which optical properties should diverge from accurate values of $C_\mathrm{ext}$, $C_\mathrm{sca}$, $C_\mathrm{abs}$ and $g$, denoted as $\lambda_\mathrm{min,\beta=1}$), obtained using the traditional numerical value of $\beta=1$ in DDA validity criteria (i) (Eq.~\ref{eq:validity_criteria_i}). We also list the actual wavelength where this occurs for this specific set of particle structures/properties ($\lambda_\mathrm{div}$). This wavelength is defined as the point where the optical cross-section diverges (by more than 10\%) from the benchmark model (highest number of dipoles) for each shape type and dipole number, with all values obtained at $R_\mathrm{mie}=0.5~\upmu$m. The (log-wavelength) average percent error versus the benchmark values, at all wavelengths above ($\lambda_\mathrm{div}$), is listed also. The largest dipole number in each case always satisfies the validity criteria, and is our benchmark model, and thus does not have a divergence wavelength. There is no table for $R_\mathrm{mie}=0.1~\upmu$m because the results for DDA do not diverge by more than 10\% at any point for any shape/dipole number studied (though the results for MMF and Mie theories do).}
	\label{table:Beta_and_lambda_min_values}
    \setlength{\tabcolsep}{3.5pt} % reduce white space to fit table onto one page
    \begin{tabular}{*{15}{cccccrccrccrccr}} % NOTE: if we ever need more than 10 columns, just add the number in curly braces! e.g. the {12} here.
    \hline
    \multicolumn{3}{c}{} &  \multicolumn{3}{c}{$C_\mathrm{ext}$} & \multicolumn{3}{c}{$C_\mathrm{sca}$} & \multicolumn{3}{c}{$C_\mathrm{abs}$} & \multicolumn{3}{c}{$g$}\\
    \cmidrule(lr){4-6} \cmidrule(lr){7-9} \cmidrule(lr){10-12} \cmidrule(lr){13-15}
    Shape type   & Dipole Number $N$ & $\lambda_\mathrm{min,\beta=1}$ & $\lambda_\mathrm{div}$ & $\beta$ & $\%~\mathrm{err.}$ & $\lambda_\mathrm{div}$ & $\beta$ & $\%~\mathrm{err.}$ & $\lambda_\mathrm{div}$ & $\beta$ & $\%~\mathrm{err.}$ & $\lambda_\mathrm{div}$ & $\beta$ & $\%~\mathrm{err.}$\\
    \hline
                          & 131    & 1.918 & 0.497 & 3.365 & 7.6\% & 0.307 & 4.555 & 8.0\% & 0.560 & 3.021 & 8.2\% & 0.800 & 2.108 & 7.3\% \\
                          & 1085   & 0.837 &   -   &   -   & 4.1\% &   -   &   -   & 5.2\% & 0.258 & 2.451 & 4.4\% & 0.253 & 2.426 & 4.1\% \\
                          & 9410   & 0.380 &   -   &   -   & 1.6\% &   -   &   -   & 2.1\% &   -   &   -   & 1.7\% & -     & -     & 1.7\% \\   
    Linear branched       & 81313  &   -   &   -   &   -   &   -    &   -   &   -   &   -    &   -   &   -   &   -    &   -   &   -   &   -    \\
    \cmidrule(lr){2-15}
                          & Mie    &       &       &       & 36.9\% &       &       & 54.3\% &       &       & 42.1\% &       &       & 34.3\%\\
                          & MMF    &       &       &       & 24.2\% &       &       & 27.4\% &       &       & 23.8\% &       &       & 4.5\%\\
    \hline
                          & 29     & 3.924 & 0.879 & 3.197 & 3.3\% & 0.507 & 5.471 & 4.0\% & 1.020 & 2.794 & 3.2\% & 1.380 & 2.156 & 3.7\% \\  
                          & 233    & 1.509 & 0.354 & 3.525 & 1.1\% &   -   &   -   & 2.4\% & 0.482 & 2.849 & 1.3\% & 0.490 & 2.809 & 2.3\% \\  
                          & 1787   & 0.701 &   -   &   -   & 2.0\% &   -   &   -   & 2.8\% &   -   &   -   & 2.4\% &   -   &   -   & 2.9\% \\  
    Elongated cluster     & 15604  &   -   &   -   &   -   & 0.9\% &   -   &   -   & 0.9\% &   -   &   -   & 0.8\% &   -   &   -   & 0.9\% \\
                          & 131130 &   -   &   -   &   -   &   -   &   -   &   -   &   -   &   -   &   -   &   -   &   -   &   -   &   -   \\
    \cmidrule(lr){2-15}
                          &  Mie   &       &       &       & 27.8\% &       &       & 25.1\% &       &       & 30.8\% &       &       & 19.7\% \\
                          &  MMF   &       &       &       & 18.8\% &       &       & 23.8\% &       &       & 24.9\% &       &       & 6.7\% \\
    \hline
                          & 90     & 2.261 & 0.593 & 3.227 & 7.3\% & 0.295 & 5.275 & 4.1\% & 0.685 & 2.770 & 8.2\% & 0.750 & 2.536 & 3.6\% \\
                          & 675    & 0.996 &   -   &   -   & 4.0\% &   -   &   -   & 2.1\% & 0.320 & 2.587 & 4.5\% & 0.284 & 2.785 & 0.8\% \\
                          & 5043   & 0.496 &   -   &   -   & 1.2\% &   -   &   -   & 1.7\% &   -   &   -   & 1.2\% & -     &   -   & 1.7\% \\
    Compact cluster       & 43730  &   -   &   -   &   -   &   -   &   -   &   -   &   -   &   -   &   -   &   -   & -     &    -  &   -   \\  
    \cmidrule(lr){2-15}
                          &   Mie  &       &       &       & 24.6\% &       &       & 24.7\% &       &       & 26.1\% &       &       & 21.9\% \\
                          &   MMF  &       &       &       & 26.3\% &       &       & 22.0\% &       &       & 19.8\% &       &       & 15.4\%\\
    \hline
    \end{tabular}
\end{table*}

We are not the first to notice that the numerical value of $\beta$ may be able to be relaxed for some specific parameter spaces \citep{Zubko_Petrov_Grynko_Shkuratov_Okamoto_Muinonen_Nousiainen_Kimura_Yamamoto_Videen_2010}, but here we wish to apply the same idea to a comprehensive study of atmospheric fractals. However, any tentative assumption that the validity criteria be relaxed should be explored with a variety of refractive index profiles and a wide range of particles sizes before it can be applied generally with confidence. For demonstrative purposes we use $\beta=2$ for the rest of this paper. While this extends to shorter wavelengths than previously assumed, this still errs on the side of caution compared to most of the values obtained in Table \ref{table:Beta_and_lambda_min_values}. At longer wavelengths, there is some divergence in the DDA models -- in particular, that the lower resolution models predict larger cross-sections by up to 20\% (see residuals in Appendix~\ref{appendix:residuals}). However, these uncertainties are still much smaller than those of Mie and MMF at the same wavelengths.

\subsection{Asymmetry parameter}

\begin{figure*}
    \includegraphics[width=\textwidth]{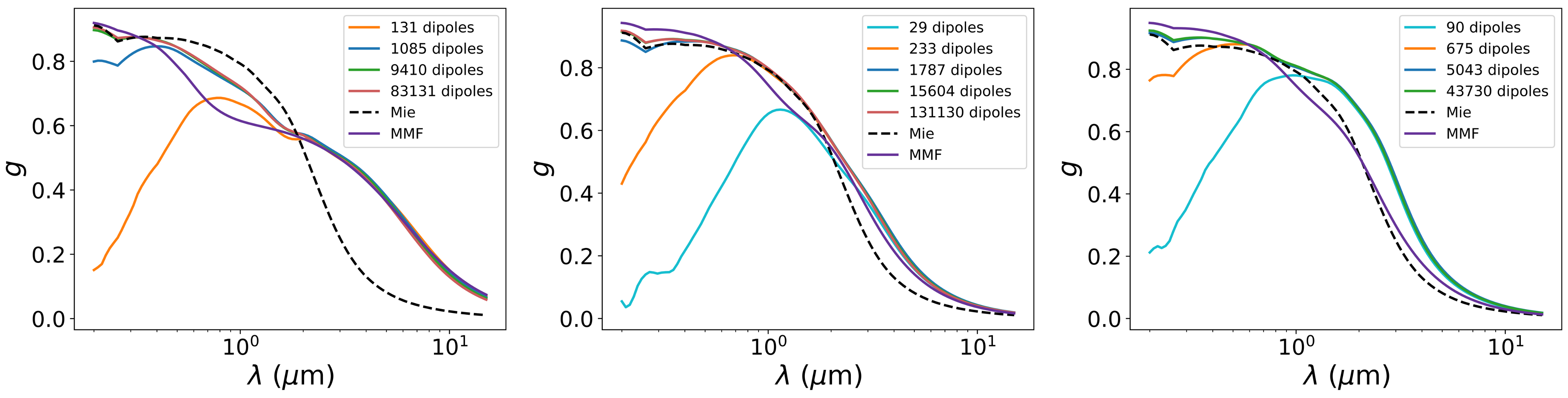}
    \caption{The asymmetry parameter $g=\braket{\cos\theta}$, plotted for each resolution of the 0.5~$\upmu$m linear branched, elongated cluster, and compact cluster particles (from left to right). Residuals are plotted in Fig.~\ref{fig:residuals_g_asymmetry}.}
    \label{fig:results_g_asymmetry}
\end{figure*}

In Fig.~\ref{fig:results_g_asymmetry}, we plot the asymmetry parameter for each shape. Similar to the cross-sections, the low-resolution models reproduce the correct value very well for longer wavelengths, and there is a divergence wavelength below which the results are invalid. This wavelength ($\lambda_\mathrm{div}$) is calculated and tabulated for each shape type/resolution in Table~\ref{table:Beta_and_lambda_min_values}. The result for the compact cluster ($d_{f}=2.529$) is especially significant. MMF has been shown to be less successful at reproducing the angle-dependent scattering properties of particles with $d_f>2$ \citep{tazaki2016light}. Our results here are consistent with this; in Table \ref{table:Beta_and_lambda_min_values} the average percent error for MMF in $g$ is shown to be only 4.5\% for the linear branched particle ($d_f=1.811$), but then 15.4\% for the compact cluster ($d_{f}=2.529$). In comparison, DDA is much more successful, even at the lowest resolutions. The lowest resolution (90 dipole) DDA model of the compact cluster reproduced $g$ with an average error of only 3.6\% (for values above the divergence wavelength of $0.750~\upmu$m), presumably because the structure of it's dipoles still estimate the shape of the particle reasonably well. It will be of great use to the astrophysical community to be able to make fast and accurate calculations of the asymmetry parameter; for guidance on estimated errors at each resolution, the differences are highlighted more clearly in the residuals (Fig.~\ref{fig:residuals_g_asymmetry}).

\subsection{DDA - Mapping the parameter space of validity criteria (i)}

\begin{figure}
    \includegraphics[width=\columnwidth]{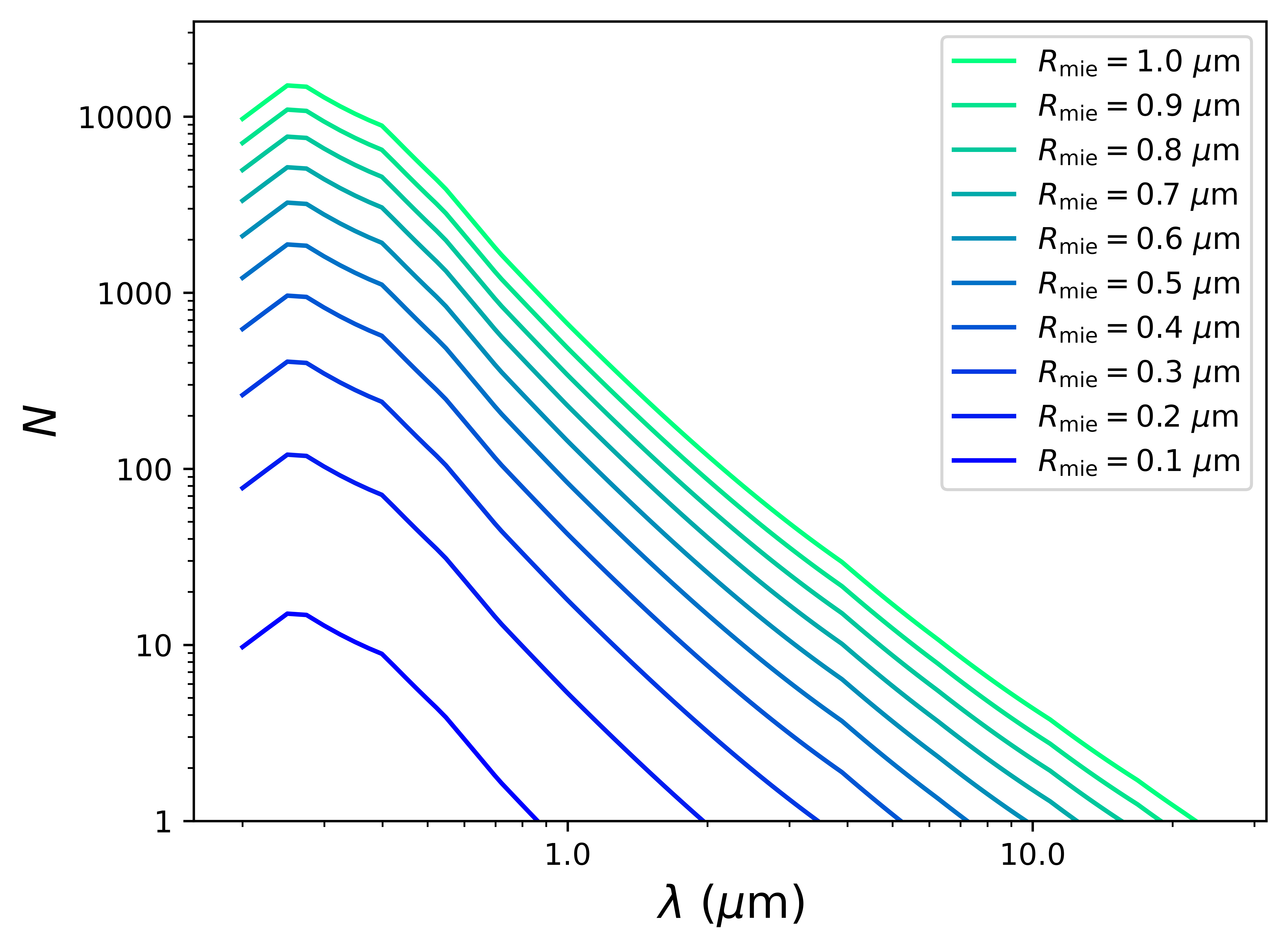}
    \caption{The minimum number of dipoles needed to correctly predict optical cross-sections using DDA at various particle sizes, using Eq.~\ref{eq:N_parameter_space} and assuming a value of $\beta=2$ (determined numerically in section \ref{sec:results}). We can obtain accurate results for small particles and long wavelengths with a very small number dipoles. The largest particle sizes and shortest wavelengths require the highest dipole number (and longest computation time).}
    \label{fig:N_plotted}
\end{figure}

A key research question presented earlier was to determine the minimum number of dipoles needed to correctly predict optical properties for a particular wavelength and particle size. Starting with the original validity criteria (i):
\begin{equation} \label{eq:DDA_validty_Beta}
    |m(\lambda)|kd<\beta.
\end{equation}
We can substitute Eq.~\ref{eq:DDA_dipole_size_d} into \ref{eq:DDA_validty_Beta} and express $k$ in terms of $\lambda$:
\begin{equation}
    |m(\lambda)|\frac{2\pi}{\lambda} \sqrt[3]{\frac{4 \pi}{3N}} R_\mathrm{mie} < \beta.
\end{equation}
Finally, rearranging for number of dipoles $N$ gives:
\begin{equation} \label{eq:N_parameter_space}
    N > \frac{32}{3}\pi^{4} \left(\frac{|m(\lambda)| R_\mathrm{mie}}{\beta \lambda}\right)^{3}.
\end{equation}
Therefore, for a specific wavelength $\lambda$, particle size $R_\mathrm{mie}$, and refractive index profile $m(\lambda)$, we can calculate the minimum number of dipoles that our shape file needs to contain for DDA to be accurate, using the numerical parameter determined in section \ref{sec:results} as $\beta=2$. Eq.~\ref{eq:N_parameter_space} is plotted for several different values of $R_\mathrm{mie}$ in Figure~\ref{fig:N_plotted} to give an idea of the minimum dipole number required for particles of a specific size and at various wavelengths. For example, if it was desired to determine optical cross-sections for $R_\mathrm{mie}=0.1~\upmu$m particles at $\lambda=1~\upmu$m, the minimum number of dipoles given by Eq.~\ref{eq:N_parameter_space} is fewer than $N=1$, meaning that validity criteria (i) is satisfied for shapes of any resolution (though care should be taken to ensure that validity criteria (ii) is satisfied -- the dipole distribution should still at least vaguely represent the shape). However, if we wanted to analyse larger particles ($R_\mathrm{mie}=0.5~\upmu$m) at same wavelength, Eq.~\ref{eq:N_parameter_space} predicts that $N>83$ dipoles would be required. This agrees with the curves in Figure~\ref{fig:results_elongated}, where 29-dipole shape is beginning to diverge from the benchmark at $\lambda=1~\upmu$m, whereas all others (including the next lowest-resolution shape $N=90$) still give accurate results. 

An interesting discussion point is that the refractive index profile can affect the parameter space where it changes significantly as a function of wavelength, and as a result the number of dipoles required can actually start decreasing again as wavelengths get smaller (the curve profile inverts between $0.2-0.3~\upmu$m in Figure~\ref{fig:N_plotted}). This is as a result of the refractive index profile sharply decreasing, affecting Eq.~\ref{eq:N_parameter_space}. Flat refractive index profiles would form simple straight lines. 

When using DDA, it would always be preferable to use the smallest dipole number possible because, for each increase in $N$, the computation time increases by a factor of $\approx N^2$. Equation~\ref{eq:N_parameter_space} therefore gives a methodology and pathway for selecting a minimum number of dipoles for a specific shape-type, from which accurate optical cross-sections can be calculated to within a few percent in the fastest possible time, providing that the dipole distribution at least coarsely represents the shape. However, if this equation is used for new shape-types, it is strongly recommended to repeat the process outlined in this paper and check for convergence between models of increasing resolution before using the cross-sections obtained.

If we simplify our model by assuming a refractive index profile that is independent of $\lambda$ (for example $|m|\approx 1.6$, an approximate value for the exoplanet haze analogues investigated in \citet{he2018laboratory}), Eq.~\ref{eq:N_parameter_space} becomes:

\begin{equation} \label{eq:N_parameter_space_simplified}
    N \approx 208 \left(\frac{R_\mathrm{mie}}{\lambda}\right)^{3}.
\end{equation}
Simplifying to Eq.~\ref{eq:N_parameter_space_simplified}, although not realistic in most cases, can give insight into the simple scaling of the dipole number as $R_\mathrm{mie}$ and $\lambda$ are altered; it highlights that firstly, the criteria requires shapes on the order of a few 100 dipoles to represent particles where $R_\mathrm{mie} \approx \lambda $. It also shows that where $R_\mathrm{mie} < \lambda$, the minimum number of dipoles becomes very small and we can save a lot of computation time. Finally, it shows that we would need a much larger $N$ to represent large particles ($R_\mathrm{mie}$) at small wavelengths ($\lambda$). This remains a challenge for DDA in general; however, for the particular application of exoplanet atmospheres, very small wavelengths compared to the particle radius are infrequently modelled due to the wavelength range of the telescopes, so it may not be an issue for our specific intended application. More depth concerning the short and long wavelength ranges and applicability of DDA will be explored in a future paper (Lodge et al., in prep).

\section{Conclusions} \label{sec:conclusions}

To conclude, the main points of this paper can be summarised as follows:
\begin{enumerate}

    \item For a soot-based refractive index profile, fractal aggregates are predicted to absorb and scatter much more radiation than their spherical counterparts at almost all wavelengths, but especially at the shortest and longest wavelengths. This is in contrast to a result from \cite{wolf2010fractal}, where fractal shapes with a different refractive index profile were shown to be transmittal at long wavelengths. The refractive index profile chosen is shown to therefore have significant potential impacts for the energy balance of hazy exoplanet and brown dwarf atmospheres, where high concentrations of these particles exist.
        
    \item Where particles are expected to be more complex in geometry than spheres, and calculating accurate optical cross-sections is crucial, it is necessary to use a complex analysis. While there are advantages to using Mie and MMF models of aerosols, these methods cannot obtain the same level of accuracy as DDA, especially for the asymmetry parameter $g$. If Mie or MMF are used, it is likely that the absorptive and scattering properties of fractal aerosols will be underestimated, and any subsequent radiative analysis may propagate these underestimations as a result. In particular, these theories miss the high absorption at long wavelengths mentioned above.

    \item We show that it is possible to obtain accurate optical cross-sections ($C_\mathrm{ext}$, $C_\mathrm{abs}$, and $C_\mathrm{sca}$) and the asymmetry parameter ($g$) quickly for non-spherical particles, using DDA but at very low resolution (dipole numbers as low as 29, providing that the distribution roughly represents the shape) for a certain wavelength range.
    
    \item Following on from this, the minimum dipole number to satisfy DDA validity criteria (i) (Eq.~\ref{eq:validity_criteria_i}) for a particle particle size and wavelength is given by Equation~\ref{eq:N_parameter_space}. This could speed up the calculation dramatically (by allowing choice of the smallest possible number of dipoles), which would enable DDA to be used directly within radiative transfer models or speed up the preparation of pre-calculated optical property grids for more computationally-intensive purposes, allowing a finer-resolution parameter space to be probed.
    
    \item It is tentatively suggested that if tolerances of $\approx10\%$ are acceptable, the numerical criteria of validity for DDA may be relaxed, replacing the value $\beta=2$ in validity DDA criteria (i). However, a study of whether this remains true for a wide range of refractive index profiles and particle sizes should take place before it is applied with confidence.

\end{enumerate}

In summary, detailed analysis of aerosols is absolutely necessary, given the significant differences in the optical properties predicted by different theoretical methods, and the extreme effects that these might have on the energy balance of exoplanet and brown dwarf atmospheres. DDA is a very powerful tool that may help achieve this goal; this study could naturally be extended to consider particles made from inhomogeneous chemical compositions, as well as multiple scattering from a distribution of different shape types, to test whether low-resolution models still give accurate optical properties under different conditions. In addition, it would be of value to explore the impact of these differing models on observables -- how would the output of radiative transfer codes differ if we use the cross-sections obtained by DDA instead of Mie or MMF theory, for example? Modelling how light interacts with matter from other worlds is a fascinating field to explore, and we look forward to investigating all of these questions and ideas in more depth in future papers. 
\section*{Acknowledgements}

 ML would like to acknowledge the generous support of the Keith Burgess Scholarship that allowed this research to be carried out. The authors thank K. Adachi, P. Buseck and S. Chung for sharing their TEM shape files, as well as R. Tazaki, D. Grant, P. Carter, M. Min, É. Hébrard, M. O'Donnell and S. Moran for stimulating ideas and discussions which helped formulate the ideas presented here. We also thank an anonymous reviewer whose comments significantly enhanced this paper. Additionally, we thank the authors of \texttt{optool} \citep{2021ascl.soft04010D}, \texttt{DDSCAT} \citep{draine1994discrete}, \texttt{matplotlib} \citep{Hunter:2007} and \texttt{NumPy} \citep{harris2020array}for making their software freely available. HRW and ZML acknowledge the financial support from the Science and Technologies Facilities Council grant number ST/V000454/1. HRW was also partly funded by UK Research and Innovation (UKRI) under the UK government’s Horizon Europe funding guarantee for an ERC STG award [grant number EP/Y006313/1].This work was carried out using the computational facilities of the Advanced Computing Research Centre, University of Bristol - http://www.bristol.ac.uk/acrc/.

%%%%%%%%%%%%%%%%%%%%%%%%%%%%%%%%%%%%%%%%%%%%%%%%%%
\section*{Data Availability}

All codes used in this project are made publicly available, and testing of them is freely encouraged:
\begin{itemize}
    \item \url{https://github.com/mglodge/CORAL.git}
    \item \url{https://github.com/mglodge/STAG.git}
    \item \url{https://github.com/mglodge/SPHERIFY.git}
\end{itemize}

The v1.0 release of CORAL (used in this paper) is available at: \url{https://doi.org/10.5281/zenodo.10257651}.

In addition, the shape files and refractive index data used in the models are made available in the supplementary materials.



\bibliographystyle{mnras}
\bibliography{references} % if your bibtex file is called references.bib

%\begin{thebibliography}{99}
%\end{thebibliography}

%%%%%%%%%%%%%%%%%%%%%%%%%%%%%%%%%%%%%%%%%%%%%%%%%%

%%%%%%%%%%%%%%%%% APPENDICES %%%%%%%%%%%%%%%%%%%%%

\appendix

\section{Geometrical Cross-Section} \label{appendix:geo_cross_section}

The Geometrical cross-section of a fractal aggregate can be found using the method outlined in \citet{Tazaki_2021}:
\begin{equation}
    G=N \pi R_{0}^{2}\left\{ 
    \begin{array}{ c l }
    12.5N^{-0.315} \exp(-2.53/N^{0.0920}) & \quad \textrm{if } N < N_\mathrm{th} \\
    \frac{A}{1+(N-1)\tilde{\sigma}}                 & \quad \textrm{otherwise,}
    \end{array}
    \right. 
\end{equation}
where $\tilde{\sigma}$ is the overlapping efficiency, $A=1$ and $N_\mathrm{th}$ is given by
\begin{equation}
    N_\mathrm{th}=\min{(11D-8.5,8)}.
\end{equation}
$A$ can be calculated numerically to join the two regimes, but this can cause small errors in the short-wavelength limit because of the use of two different models that are not consistent with each other (private comms, Tazaki), and so it is tentatively given as 1 to keep consistent between the models used (consistent with implementation in \texttt{optool}). The overlapping efficiency can be calculated using the following two equations: 
\begin{equation}\label{eq:tilde_sigma}
    \tilde{\sigma} = \frac{\eta^{\frac{2}{d_{f}}}}{16} \int_{\eta}^{\infty} x^{-\frac{2}{d_{f}}}e^{-x} dx,
\end{equation}
\begin{equation} \label{eq:eta}
    \eta = 2^{d_{f}-1} \frac{k_{0}}{N}.
\end{equation}
Note that \citet{Tazaki_2021} provides a much more detailed description of these terms, derivations and justification for their use, as well as (very helpfully) more efficient numerical methods of calculating them than listed here; we only wish to outline the procedure for absolute clarity of the methods that have been used for this particular research, and direct the reader to this original paper if they would like to know more.

\section{Full Expansion of DDA Equation}\label{appendix:expanded_DDA_matrix}

This section aims to clarify the construction of the $\mathbf{A_{jk}}$ tensor described by Eq. \ref{A_jk_equation}. Considering a very small system of $N=3$ dipoles $j=1,2,3$, made of the same material, Eq. \ref{compact_linear} for the three dipoles $j$ becomes:
\begin{align}
    \sum_{k=1}^{N}\mathbf{A_{jk}P_{k}}&=\mathbf{E_{inc,j}}, \\
    \mathbf{ A_{11}P_{1} + A_{12}P_{2} + A_{13}P_{3} }&=\mathbf{E_{inc,1}}, \\
\mathbf{ A_{21}P_{1} + A_{22}P_{2} + A_{23}P_{3} }&=\mathbf{E_{inc,2}}, \\
\mathbf{ A_{31}P_{1} + A_{32}P_{2} + A_{33}P_{3} }&=\mathbf{E_{inc,3}}.
\end{align}
Each line of the equation describes the resultant electric field at $j$ as a result of the polarisability of the dipole $j$ (through the $\mathbf{A_{jj}}$ term, e.g. $\mathbf{A_{11}}$), and the polarisation and position of the other dipoles (through the $\mathbf{A_{jk}}$ terms, for example $\mathbf{A_{12}}$ and $\mathbf{A_{13}}$). In compact matrix form this can be written:
\begin{equation}
    \begin{pmatrix}
        \mathbf{A_{11}} & \mathbf{A_{12}} & \mathbf{A_{13}}  \\
        \mathbf{A_{21}} & \mathbf{A_{22}} & \mathbf{A_{23}}  \\
        \mathbf{A_{31}} & \mathbf{A_{32}} & \mathbf{A_{33}} 
    \end{pmatrix}
    \begin{pmatrix}
        \mathbf{P_{1}} \\
        \mathbf{P_{2}} \\
        \mathbf{P_{3}}
    \end{pmatrix}
    =
    \begin{pmatrix}
        \mathbf{E_{inc,1}} \\
        \mathbf{E_{inc,2}} \\
        \mathbf{E_{inc,3}}
    \end{pmatrix}.
\end{equation}
Note that $\mathbf{P}$ and $\mathbf{E_{inc}}$ are 3-dimensional vectors, and the $\mathbf{A_{jk}}$ elements are each a $3\times3$ matrix. Substituting in $\mathbf{A_{jj}}=\alpha^{-1}$, applying the delta function in Eq. \ref{alpha_cldr}, and expanding to show the individual cartesian elements of each vector, we construct the matrix equation:

\newpage

\begin{strip}
\begin{equation} \label{linear_expanded}
    \begin{pmatrix}
        \alpha^{-1}_{xx} & 0 & 0 & A_{12,xx} & A_{12,xy} & A_{12,xz} & A_{13,xx} & A_{13,xy} & A_{13,xz} \\
        0 & \alpha^{-1}_{yy} & 0 & A_{12,yx} & A_{12,yy} & A_{12,yz} & A_{13,yx} & A_{13,yy} & A_{13,yz} \\
        0 & 0 & \alpha^{-1}_{zz} & A_{12,zx} & A_{12,zy} & A_{12,zz} & A_{13,zx} & A_{13,zy} & A_{13,zz} \\
        
        A_{21,xx} & A_{21,xy} & A_{21,xz} & \alpha^{-1}_{xx} & 0 & 0 & A_{23,xx} & A_{23,xy} & A_{23,xz} \\
        A_{21,yx} & A_{21,yy} & A_{21,yz} & 0 & \alpha^{-1}_{yy} & 0 & A_{23,yx} & A_{23,yy} & A_{23,yz} \\
        A_{21,zx} & A_{21,zy} & A_{21,zz} & 0 & 0 & \alpha^{-1}_{zz} & A_{23,zx} & A_{23,zy} & A_{23,zz} \\
        
        A_{31,xx} & A_{31,xy} & A_{31,xz} & A_{32,xx} & A_{32,xy} & A_{32,xz} & \alpha^{-1}_{xx} & 0 & 0 \\
        A_{31,yx} & A_{31,yy} & A_{31,yz} & A_{32,yx} & A_{32,yy} & A_{32,yz} & 0 & \alpha^{-1}_{yy} & 0 \\
        A_{31,zx} & A_{31,zy} & A_{31,zz} & A_{32,zx} & A_{32,zy} & A_{32,zz} & 0 & 0 & \alpha^{-1}_{zz}
    \end{pmatrix}
    \begin{pmatrix}
        P_{1,x} \\
        P_{1,y} \\
        P_{1,z} \\
        
        P_{2,x} \\
        P_{2,y} \\
        P_{2,z} \\
        
        P_{3,x} \\
        P_{3,y} \\
        P_{3,z} 
    \end{pmatrix}
    =
    \begin{pmatrix}
        E_{1,x} \\
        E_{1,y} \\
        E_{1,z} \\
        
        E_{2,x} \\
        E_{2,y} \\
        E_{2,z} \\
        
        E_{3,x} \\
        E_{3,y} \\
        E_{3,z} 
    \end{pmatrix}.
\end{equation}	
\end{strip}
The Electric field terms can be found for dipole $j$ using: 

\begin{equation}
    \mathbf{E_{inc,j}}=
    \begin{pmatrix}
        E_{j,x} \\
        E_{j,y} \\
        E_{j,z} \\
    \end{pmatrix},
\end{equation}

\begin{equation}
    \mathbf{E_{inc,j}}=
    \mathbf{\hat{e_{r}}}\exp\left(ikd\left[\cos(\phi)\cos(\theta)x-\sin(\phi)\cos(\theta)y-\sin(\theta)z\right]\right),
\end{equation}
where $\mathbf{\hat{e_{r}}}$ gives the rotated polarisation state of the Electric field (a vector, calculated using Eq.~\ref{polarisation_state_of_EM_field}), $\phi$ and $\theta$ are the angles of the incoming radiation in spherical polar coordinates, and $x$, $y$ and $z$, are the dipole coordinates, relative to each other.

We can calculate all of the $\mathbf{A_{jk}}$ terms using Eq. \ref{A_jk_equation}. So, for $N$ dipoles we need to solve a set of $3N$ equations in Eq. \ref{linear_expanded} (as shown, for 3 dipoles, this gives 9 equations to solve) to find the Polarisations $\mathbf{P}$. Once solved, we can substitute them into Eq. \ref{DDA_C_Ext} and \ref{DDA_C_abs} to find the cross-section of the shape for a given wavelength.

This example highlights the computational difficulty, given that the system above is already quite complex for an incredibly small system of 3 dipoles, whereas we need to be able to perform calculations and solve systems of equations for $10^{4}$ dipoles.

\section{Averaging Optical Properties Over a Range of Angles - DDA} \label{appendix:angle_averages}

\begin{figure}
    \includegraphics[width=\columnwidth]{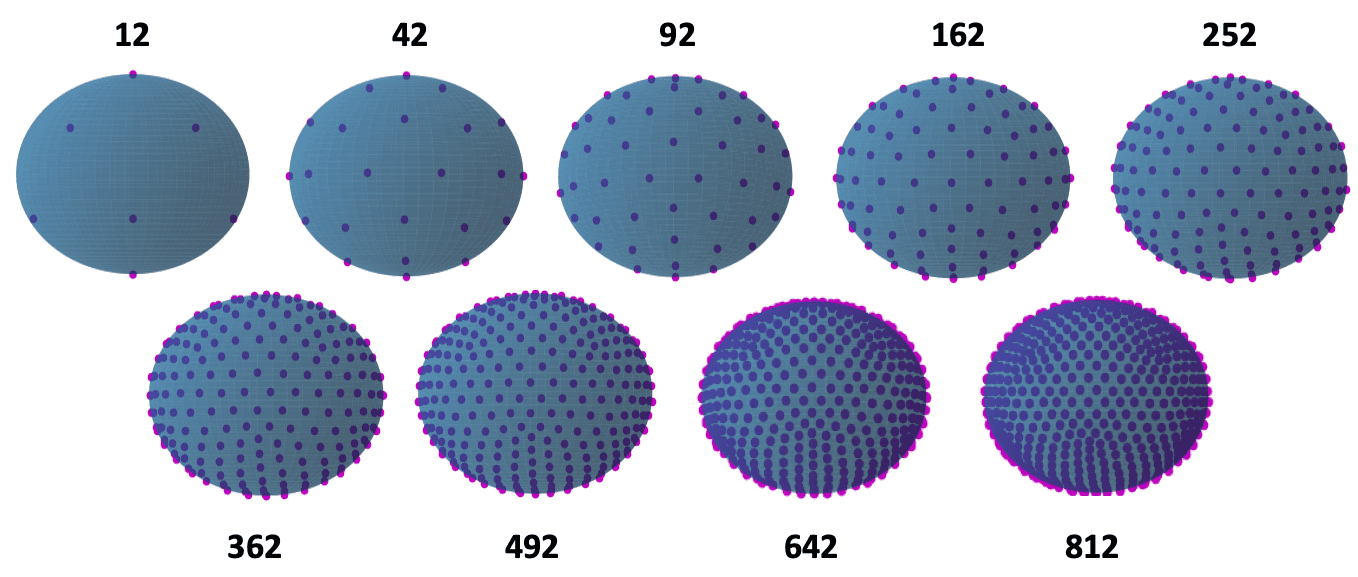}
    \caption{The distribution of angles of incoming radiation used to determine an orientational average for each shape. These positions are obtained by finding the vertices of a recursively-iterating icosahedral grid. After detailed analysis for each shape, 252 angles was chosen to be the standard number of angles used to obtain the orientational averages in this study.}
    \label{fig:angle_positions}
\end{figure}

Because DDA gives the cross-section for a specific orientation of an aerosol, unless the shape exhibits symmetry (which is not true for any of the shapes considered here), we need to find an orientational average to compare it to the other models; a suitable number of angles to average over, as well as their exact positions, is a somewhat numerically flexible choice that should be analysed for each particular set of input parameters. In this study, angles were oriented on the vertices of an icosahedral grid, that uses recursive iteration to increase the number of vertices (see Section 2.1 in \citealt{Wang_Lee_2011}). The result is an equally spaced angular distribution (between any two adjacent points across the surface of a sphere) of incoming radiation angles, which can be quickly determined by choosing the "level" of recursion (the number of iterations). The first nine levels of recursion are shown in Figure~\ref{fig:angle_positions}.

\begin{figure*}
    \includegraphics[width=\textwidth]{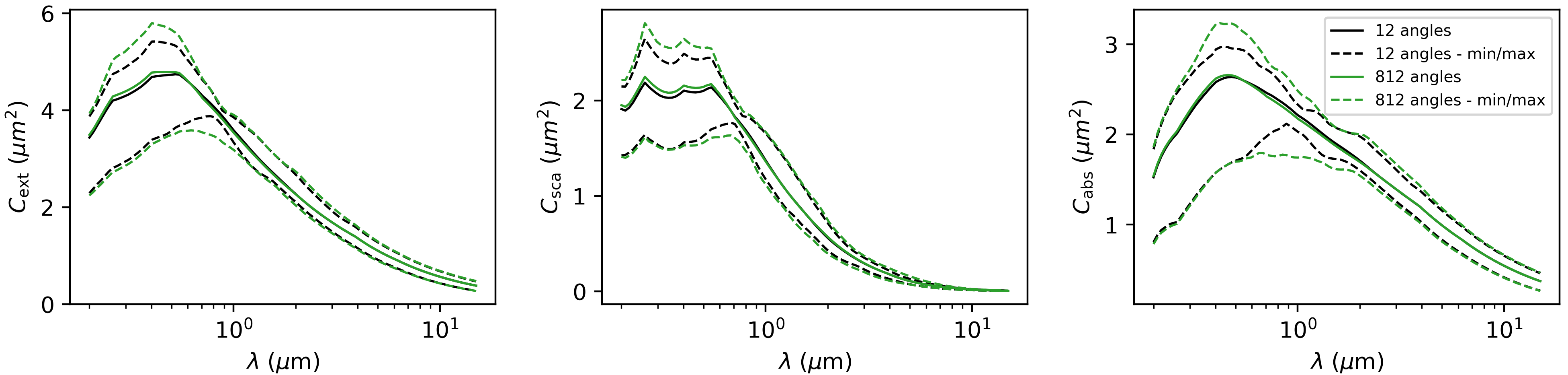}
    \caption{The black solid line indicates the orientational average cross-sections obtained for the 1085-dipole branched fractal after an analysis of 12 different angles of incoming radiation directions, whereas the dashed lines show the minimum and maximum values obtained at any one particular angle (in either polarisation state). The green lines are identical, except they represent analysis of a much higher number of angles (812) to determine the orientational average. The plot demonstrates that the range of min and max values increases when we consider more angles, but the average obtained does not change significantly, even when the number of angles is almost 2 orders of magnitude larger.}
    \label{fig:min_max_1085}
\end{figure*}

We analysed each shape file with each range of the angle distributions shown in Figure~\ref{fig:angle_positions} to determine the minimum number of angles required to obtain accurate cross-sections by comparing them to the value obtained with a very high number of angles (here an assessment with 812 different angles was chosen as the reference - this is assumed to be the `most correct' orientational average). To demonstrate the range of values that can be obtained for analysis of a single wavelength and particle shape, the min/max (of either polarisation state, whichever was smallest/largest respectively) is shown in Figure~\ref{fig:min_max_1085}.

\begin{figure*}
    \includegraphics[width=\textwidth]{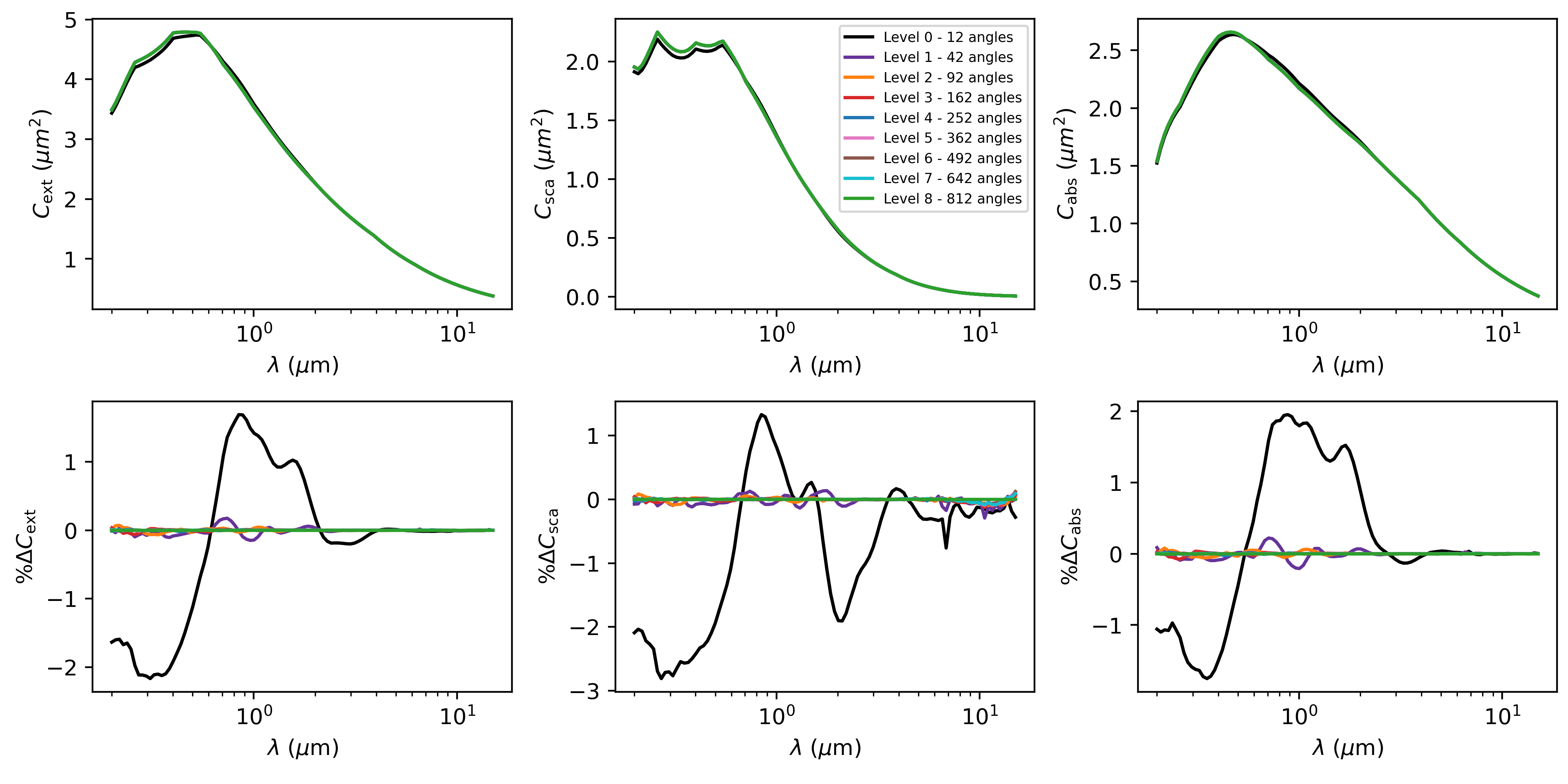}
    \caption{Top: Average orientational value of $C_\mathrm{ext}$, $C_\mathrm{sca}$ and $C_\mathrm{abs}$ for the 1085-dipole branched fractal, after averaging from different numbers of equally-spaced angles of incoming radiation. Bottom: Residual values of $C_\mathrm{ext}$, $C_\mathrm{sca}$ and $C_\mathrm{abs}$ (percent difference) versus the average found using 812 angles (our benchmark value, which considers the highest number of angles, and so is the most accurate). This figure shows that, at this resolution, calculating an orientational average using only 12 angles can retrieve a value that has (at maximum) 3\% variation compared to an analysis of 812 different angles.}
    \label{fig:angle_analysis_branched}
\end{figure*}

Figure~\ref{fig:angle_analysis_branched} goes further and compares the average values obtained from each of the incoming angle configurations in Figure~\ref{fig:angle_positions}, plotting the cross-sections obtained for the 1085-dipole branched fractal shape (at original resolution). Interestingly, the cross-sections were accurate to within 2\% even when using only 12 angles. Using 42 angles meant that the values were accurate to within 0.3\% across the entire parameter space, and similar results were found for the 1787-dipole elongated cluster shape (0.1\% max error at any point) and the 5043-dipole compact cluster (0.5\% max error).

\begin{figure*}
    \includegraphics[width=\textwidth]{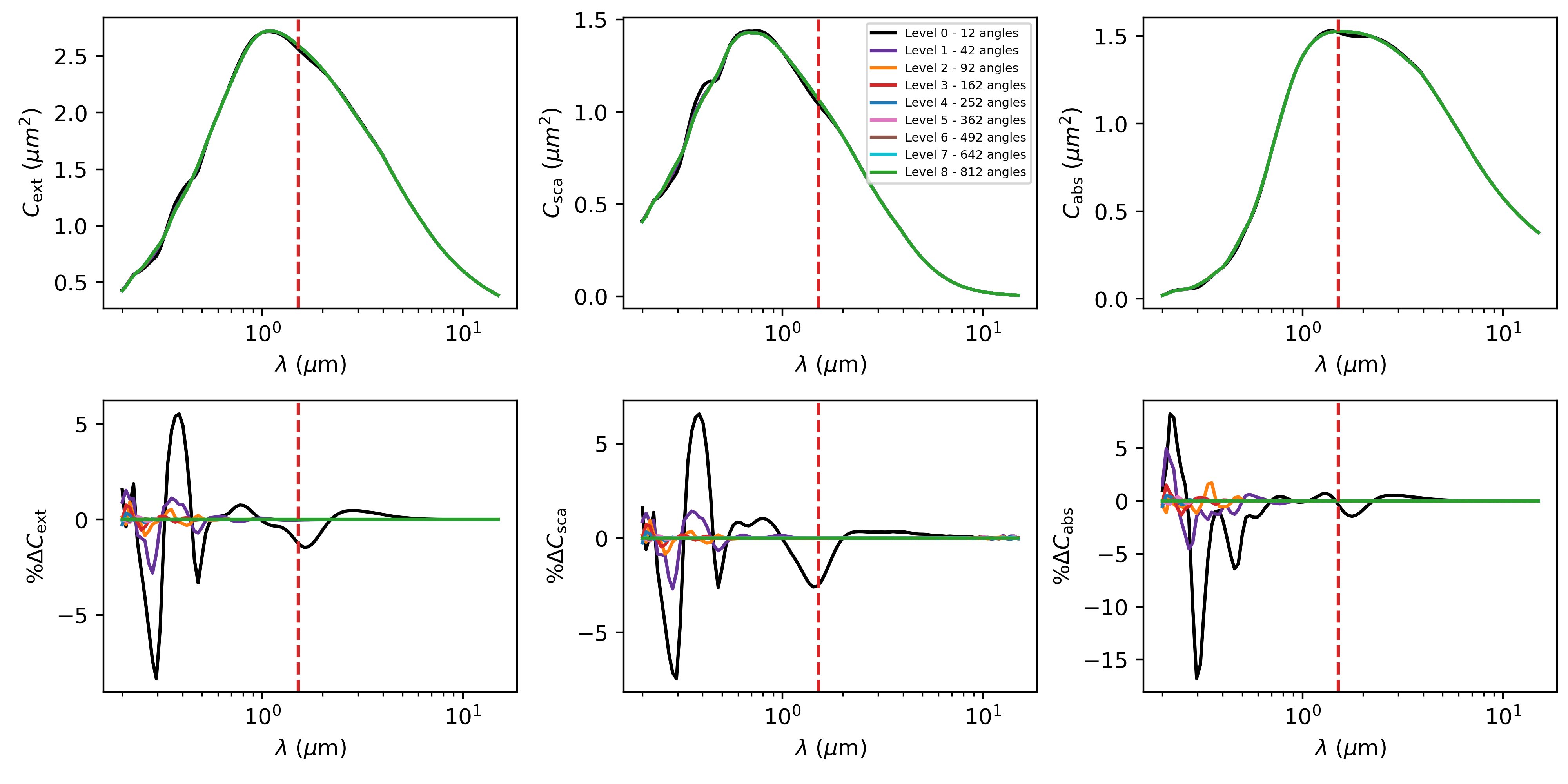}
    \caption{As in Figure~\ref{fig:angle_analysis_branched}, but for the low-resolution 29-dipole elongated cluster. Choosing fewer angles to calculate an orientational average does causes larger errors (of up to 16\%) in this case (versus a maximum of 3\% variability for the high-resolution version in Figure~\ref{fig:angle_analysis_branched}), but the largest errors only occur at the lowest wavelengths, where $\lambda<\lambda_\mathrm{min}$ (and this means the results would not be used anyway). The red dashed line marks $\lambda_\mathrm{min}$ for $N=29$ dipoles -- this is the minimum wavelength that would satisfy Eq.~\ref{eq:N_parameter_space} for $R_\mathrm{mie}=0.5$ and $\beta=2$; all wavelengths below this point would be considered invalid and discounted anyway. All values above this wavelength maintain very low variability, regardless of the number of angles used to find an orientational average. Therefore, this shows that within the region of validity, there is very little variation in the orientational average, regardless of the number of angles used to obtain it.}
    \label{fig:angle_analysis_low_res_compact}
\end{figure*}

The lowest resolution-dipole models showed higher discrepancies when using only 12 angles of up to 8\% for the 131-dipole branched fractal, 17\% for the 29-dipole elongated cluster and 8\% for the 90-dipole compact cluster. Full results for the very lowest-resolution 29-dipole elongated fractal are shown in Figure~\ref{fig:angle_analysis_low_res_compact}.

Figure~\ref{fig:angle_analysis_low_res_compact} is the lowest-resolution that we would expect to use (to still be able to represent particle shape), and the most inaccurate cross-sections after analysis of 12 angles, and so is a good analysis to choose a suitable minimum number of angles for this study. Because we wanted to ensure absolute accuracy of the orientational average, even in regions where $\lambda<\lambda_\mathrm{min}$, 252 angles was chosen (the error remains within $<1\%$ vs an analysis with 812 angles). It should be noted here that $\lambda_\mathrm{min}$ is calculated using Eq.~\ref{eq:validity_criteria_with_beta} with an updated $\beta=2$ value, as numerically determined in this paper. However, it is worth noting that the largest errors for the low-resolution shapes are all in regions where $\lambda<\lambda_\mathrm{min}$ (below the red dotted line on graph), indicating that actually, averages from as little as 12 angles could be enough to gain a very good orientational average value (within 2\% vs an analysis with 812 angles) for the parameter space where we would actually use them. This would save a significant amount of computation time (the total analysis time is directly proportional to the number of angles assessed when finding averages). Further analysis will be completed on this in a future paper; however 252 angles has been used to determine the orientational average in this paper to maintain absolute confidence in the analysis at all wavelengths, even in the region far below the expected region of validity ($\lambda < \lambda_\mathrm{min}$).

\section{Averaging Optical Properties over Polarisation states - DDA} \label{appendix:polarisation_averages}

\begin{figure*}
    \includegraphics[width=\textwidth]{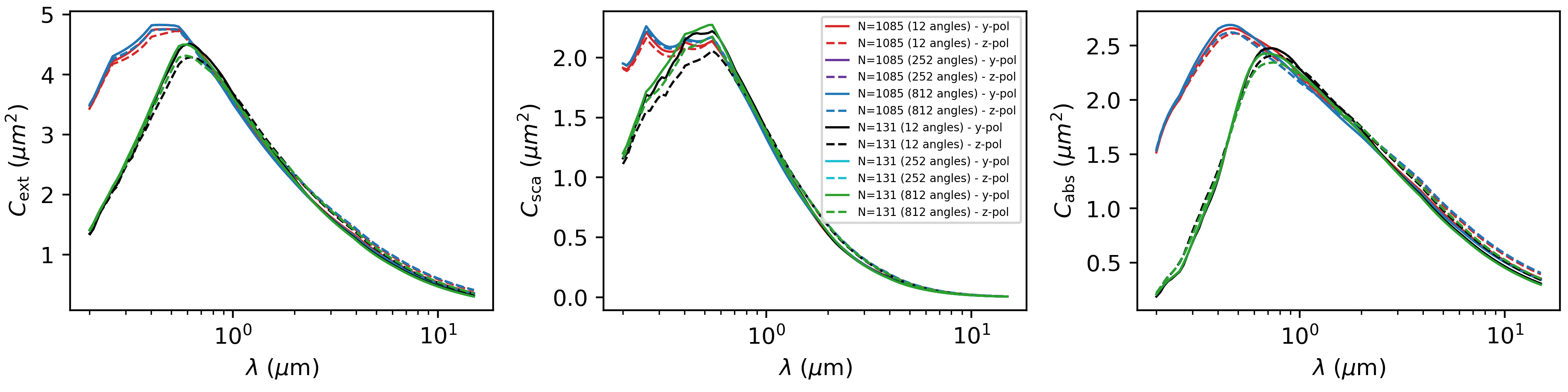}
    \caption{The solid curves indicate y-polarisation and the dashed curves represent z-polarisation states for the 1085 and 131-dipole branched fractal. There is an average difference of $\approx$6\%, 9\% and 6\% (and a maximum of 17\% difference at any one point) for $C_\mathrm{ext}$, $C_\mathrm{sca}$ and $C_\mathrm{abs}$ respectively between the $y$ and $z$ polarisation states for this shape type, regardless of the dipole number or number of angles used (shown here for 12, 252 and 812 angles). The profile for 252 angles (the number used for all figures/data published in this study) is plotted in cyan and purple, but is so close to the curve for 812 angles that they are indistinguishable (there is $<-0.1\%$ difference between the two curves on average). This clear indicates that 252 angles is more than enough to obtain accurate polarisation averages.}
    \label{fig:polarisation_analysis}
\end{figure*}

Different polarisation states have different optical cross-sections, and so most commonly, an average is taken of two perpendicular states (for radiation travelling in the x-direction, an electric field can oscillate in any position in the y-z plane; this is prior to rotation of the incoming radiation beam along, for example, one of the angles in Figure~\ref{fig:angle_positions}). Figure~\ref{fig:polarisation_analysis} shows the subtle differences obtained by individually plotting the cross-sections for an electric field oscillating in the $y$ or $z$ direction ('s' or 'p' polarisation). Because this study considers low-resolution models as well, the 131-dipole version is also shown in Figure~\ref{fig:polarisation_analysis}. In summary, the difference between polarisation states for each model is generally quite small, but not negligible; it is important to consider (and DDA is the only model to treat them separately), but using an simple average of the y and z states for each angle assessed does give a good representation of the cross-section, even for low-resolution shapes. To confirm this conclusion was equally true for the other two shape types used in this study, this analysis was repeated, with identical outcomes.

\section{Spherify results - Crosses and Cubes} \label{appendix:spherify_results}

\begin{figure*}
    \includegraphics[width=\textwidth]{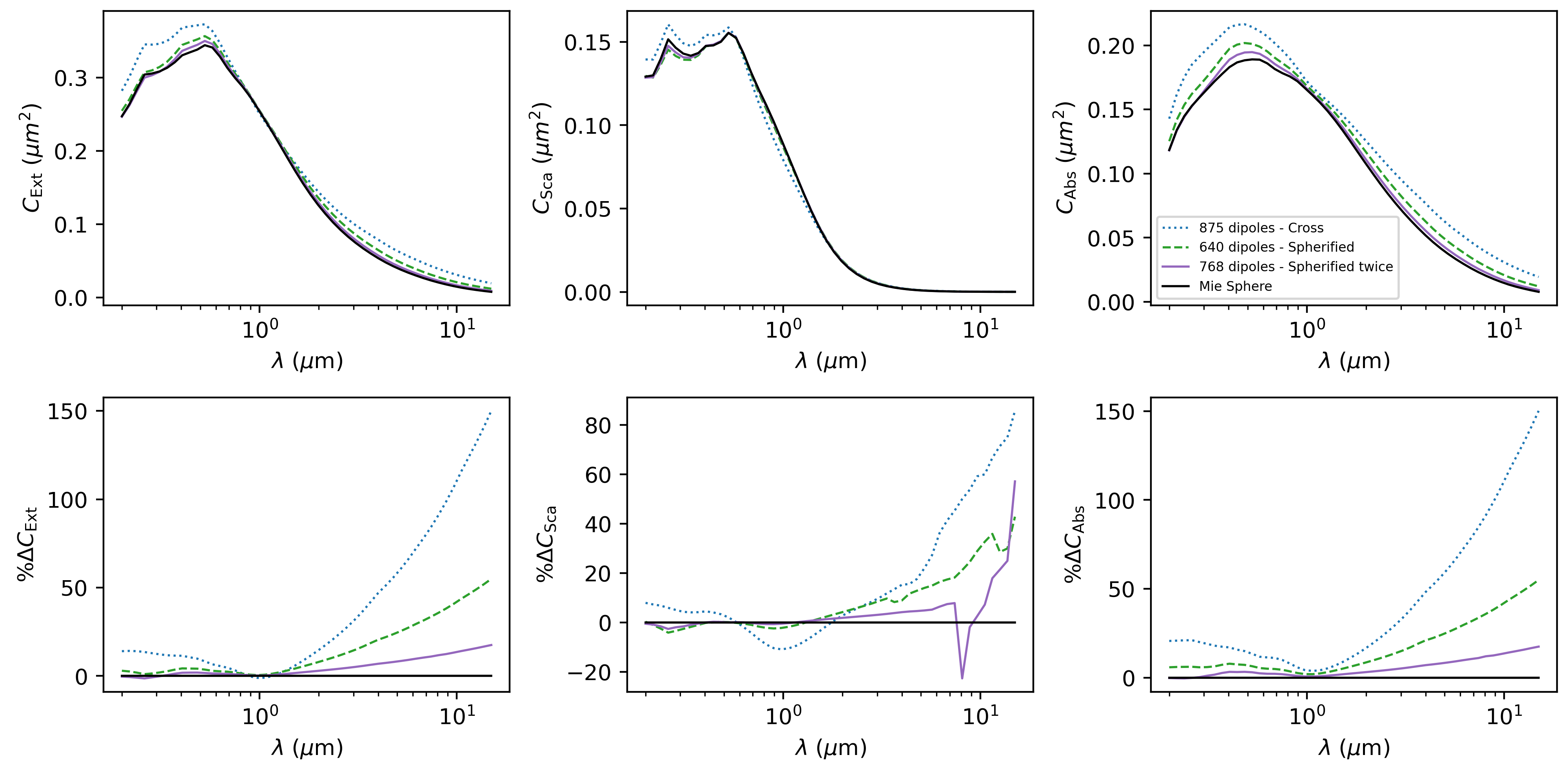}
    \caption{Scattering, absorption and extinction cross-sections for a cross that becomes more round after two iterations of the \texttt{SPHERIFY} code. Subsequent iterations become much more round (both by eye, as in Figure~\ref{fig:spherify_proof}, and by optical profile, as shown here).}
    \label{fig:spherify_cross_full}
\end{figure*}

\begin{figure*}
    \includegraphics[width=\textwidth]{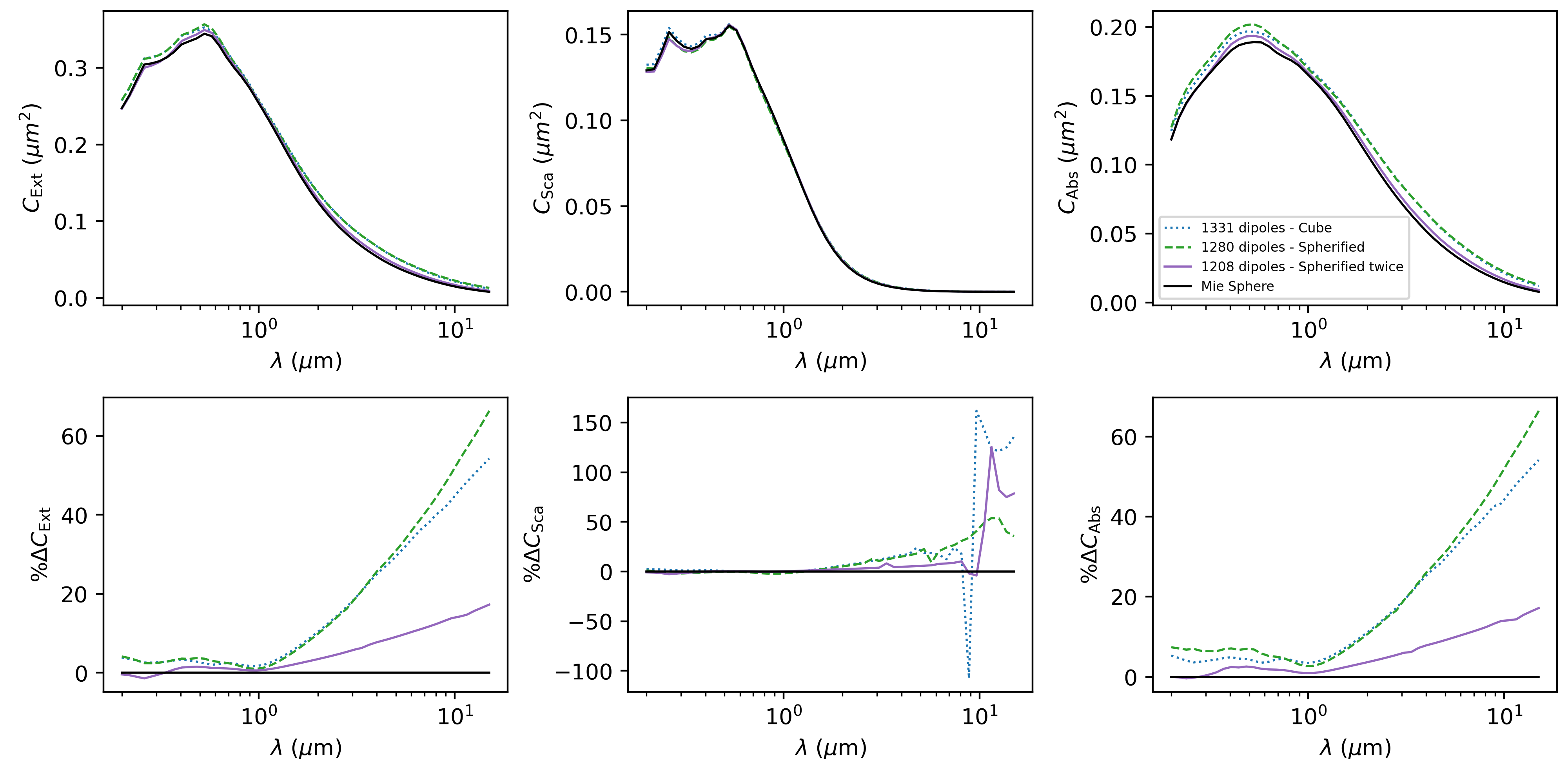}
    \caption{Scattering, absorption and extinction cross-sections for a cube that becomes more round after two iterations of the \texttt{SPHERIFY} code. The first shape formed actually has a profile further from a sphere than a cube (seen by eye in Figure~\ref{fig:spherify_proof}, the first iteration does not produce a particularly spherical shape in this case). However, a second iteration is much more round (both by eye, and by optical profile), and so the profile much more closely resembles that of a sphere.}
    \label{fig:spherify_cube_full}
\end{figure*}

Figures \ref{fig:spherify_cross_full} and \ref{fig:spherify_cube_full} show the full set of optical cross-sections after each iteration of the \texttt{SPHERIFY} code, demonstrating that both shapes become more ``round'' at the edges.

\section{Wolf and Toon (2010) - Revisited} \label{appendix:wolf_and_toon_update}

\begin{figure*}
    \includegraphics[width=\textwidth]{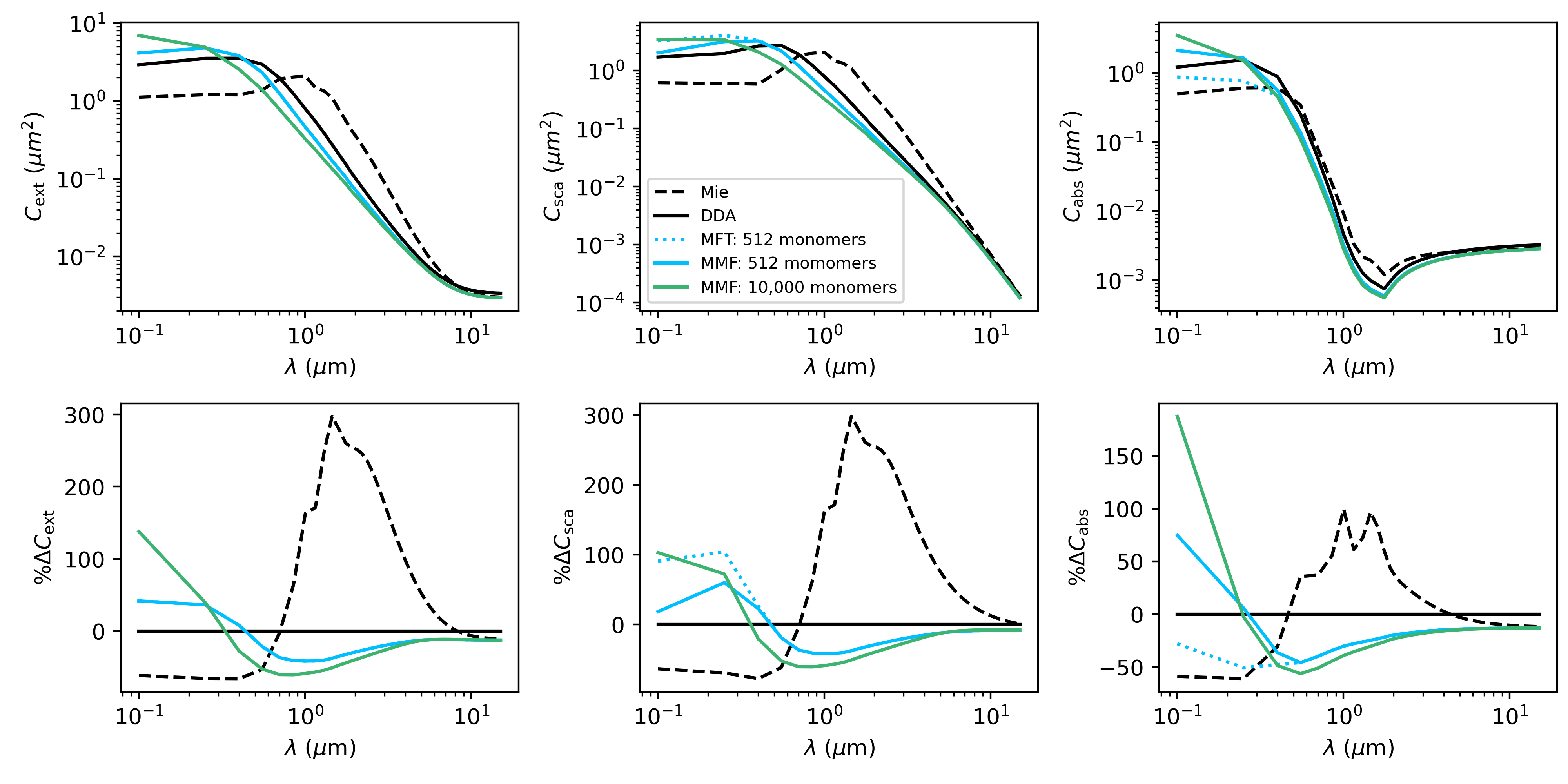}
    \caption{Top row: Extinction, scattering and absorption cross-sections for a 0.4~$\upmu$m aerosol particle, with the same MMF parameters as the linear branched fractal in Table~\ref{table:fractal_parameters}. Bottom row: The same values, but as percentage residuals versus DDA (our benchmark value).}
    \label{fig:Wolf_and_Toon_update}
\end{figure*}

Here we have aimed to reproduce Figure 3 from \citet{wolf2010fractal}, but extending to much longer wavelengths than 2~$\upmu$m, using the original refractive index data \citep{khare1984optical} and a shape type that is close to matching their original mean-field parameters of $d_\mathrm{f}=2$ and $k_0=1$ (see linear branched shape, Table \ref{table:fractal_parameters}). The number of monomers was deduced using conservation of volume (see Equation \ref{Eq:radius_of_monomer}), noting that the original paper used $R_0=50$ nm monomers, so that we would require $N_\mathrm{mon}=512$ to have the same overall volume as a 0.4~$\upmu$m radius sphere (our Mie comparison). 

The results (shown in Figure~\ref{fig:Wolf_and_Toon_update}) are in strong agreement with the original findings of \citet{wolf2010fractal}. Interestingly however, the residuals in Figure~\ref{fig:Wolf_and_Toon_update} show that the extinction of fractal aggregates (DDA) does eventually become larger than for their spherical equivalents (Mie). The cross-sections are so small at this point though, that this is likely to have negligible impact (note the logarithmic scale for the values in the top row of the figure). 

It is also interesting to note that the mean-field theory used in \citet{wolf2010fractal} has been improved (see MMF, Equation \ref{C_abs_MMF}). Here the modifications of \citet{Tazaki_Tanaka_2018} indicate that a higher proportion of light at lower wavelengths is absorbed rather than scattered, compared to the original MFT theory (Figure~\ref{fig:Wolf_and_Toon_update}). The total extinction values are unmodified (the modification only applies to $C_\mathrm{abs}$ and $C_\mathrm{sca}$).

We also repeat the MMF analysis, but forcing a higher number of monomers, in a similar manner to section \ref{results:wolf_and_toon} (we increase $N_\mathrm{mon}=10,000$, which each have a radius of $R_0=18$ nm; the supplementary notes of \citet{wolf2010fractal} suggest that anything between 10-100 nm is realistic for their model, so these are reasonable physical parameters to expect in an Archean Earth environment). Plotted in Figure~\ref{fig:Wolf_and_Toon_update}, this produces a similar conclusion to the discussion in \ref{results:wolf_and_toon}: assuming higher monomer numbers causes more extinction at short wavelengths and less at long wavelengths (a steeper profile). However, the 511 monomer model is assumed to be the more physical assumption, and more closely matches our benchmark model (DDA), so is assumed to be more accurate.

This demonstrates the need to carefully consider the correct monomer number when using MMF.

\section{Benchmark Tests} \label{appendix:benchmark_tests}

All codes have been thoroughly tested against a wide range of particle sizes, shapes, refractive indices and wavelengths of radiation. Here we perform some standard tests to compare the output against other popular codes.

\subsection{Mie}

We benchmark \texttt{CORAL} against the online Mie scattering calculator \texttt{miecalc} \citep{Prahl_mie_calc}; in Table~\ref{table:Mie benchmark}, we give values obtained by assuming that light of wavelength $\lambda=0.8~\upmu$m is incident on a sphere of radius $R_\mathrm{mie}=0.5$ with a refractive index $m=2+i$. The results agree to at least 5 significant figures.

\begin{table}
	\centering
	\caption{Table comparing the four various key output values between \texttt{miecalc} and \texttt{CORAL}.}
	\label{table:Mie benchmark}
	\begin{tabular}{ccccc} % five columns, alignment for each
		\hline
		    & $C_\mathrm{ext}~(\upmu$m) & $C_\mathrm{sca}~(\upmu$m) & $C_\mathrm{abs}~(\upmu$m) & $g$ \\
		\hline
		miecalc & 2.13000 & 1.0858 & 1.0442 & 0.75993 \\
            CORAL   & 2.13000 & 1.0858 & 1.0442 & 0.75993 \\
		\hline
	\end{tabular}
\end{table}

\subsection{MMF}

We benchmark \texttt{CORAL} against \texttt{OPTOOL}, performing the same test as in Appendix A of \citet{Tazaki_Tanaka_2018}. For a wavelength of $\lambda=0.8~\upmu$m, incident on a fractal aggregate of $N_\mathrm{mon}=64$ monomers, each with $R_0=0.5~\upmu$m, and with a refractive index $m=1.4+0.0001i$, arranged into a shape with a fractal dimension $d_f=2.0$ and a fractal prefactor $k_0=0.825$, we obtain the following mean-field terms in Table~\ref{table:MMF benchmark_d_terms} (first five orders shown only). Comparing to the values in Table A1 of \citet{Tazaki_Tanaka_2018}, these agree very well. Going further, we compare values for the actual cross sections and asymmetry parameter obtained in the two codes in Table~\ref{table:MMF benchmark_C_and_g}. The codes are in very good agreement.

\begin{table}
	\centering
	\caption{Table showing the first five terms of the mean-field coefficients to benchmark against \citep{Tazaki_Tanaka_2018}.}
	\label{table:MMF benchmark_d_terms}
	\begin{tabular}{cccc} % four columns, alignment for each
		\hline
		      & n & $\bar{d}_{1,n}^{(1)}$ & $\bar{d}_{1,n}^{(2)}$ \\
		\hline
		\citet{Tazaki_Tanaka_2018} & 1 & 0.352 + 0.245 i & 0.397 + 0.124 i \\
            CORAL                      & 1 & 0.352 + 0.245 i & 0.397 + 0.124 i \\
		\hline \\
            \citet{Tazaki_Tanaka_2018} & 2 & 0.419 + 0.119 i & 0.408 + 0.179 i \\
            CORAL                      & 2 & 0.419 + 0.119 i & 0.409 + 0.179 i \\
		\hline \\
            \citet{Tazaki_Tanaka_2018} & 3 & 0.388 - 0.039 i & 0.448 + 0.093 i \\
            CORAL                      & 3 & 0.388 - 0.039 i & 0.448 + 0.093 i \\
		\hline \\
            \citet{Tazaki_Tanaka_2018} & 4 & 0.123 - 0.111 i & 0.067 - 0.079 i \\
            CORAL                      & 4 & 0.123 - 0.111 i & 0.067 - 0.079 i \\
		\hline \\
            \citet{Tazaki_Tanaka_2018} & 5 & 0.014 - 0.024 i & 0.005 - 0.009 i \\
            CORAL                      & 5 & 0.014 - 0.024 i & 0.005 - 0.009 i \\
		\hline
	\end{tabular}
\end{table}

\begin{table}
	\centering
	\caption{Table comparing the four various key output values between \texttt{OPTOOL} and \texttt{CORAL}.}
	\label{table:MMF benchmark_C_and_g}
	\begin{tabular}{ccccc} % five columns, alignment for each
		\hline
		    & $C_\mathrm{ext}~(\upmu$m) & $C_\mathrm{sca}~(\upmu$m) & $C_\mathrm{abs}~(\upmu$m) & $g$ \\
		\hline
		OPTOOL & 92.374 & 92.285 & 0.0898 & 0.9008 \\
            CORAL  & 92.382 & 92.292 & 0.0897 & 0.9008 \\
		\hline
	\end{tabular}
\end{table}

\subsection{DDA}

We benchmark \texttt{CORAL} against \texttt{DDSCAT} for a range of wavelengths in Table~\ref{table:DDA_benchmark}. We assume that radiation is incident on the linear branched fractal aggregate of $N=1085$ dipoles, with an equivalent mie radius of $R_\mathrm{mie}=0.5~\upmu$m, and with a refractive index $m=1+0.5i$. The results shown are both orientational and polarisation averages (of the two default two perpendicular states). The codes are in very good agreement, and agree to 5 significant figures when incoming radiation is fired along individual single angles. Any small deviation in the orientational averaging is due to the slightly different ways in which they choose angles. To obtain these values, all \texttt{DDSCAT} parameters were set to default, with 'GKDLDR'=CALPHA*6, ETASCA=0.5, and angle ranges set to $0\leq \beta \leq 360$, $0\leq \theta \leq180$, $0 \leq \phi \leq 360$, each with 5 angle iterations. \texttt{CORAL} was run for orientational average level=3 (162 angles) and both code had accuracy $=10^{-5}$.

\setlength{\tabcolsep}{3.5pt} % reduce white space to fit table onto one page
\begin{table}
	\centering
	\caption{Table comparing the four various key output values between \texttt{DDSCAT} and \texttt{CORAL} for the $N=1085$ dipole linear branched fractal assessed in this study. All values are orientational averages and averages of two perpendicular polarisation states.}
	\label{table:DDA_benchmark}
	\begin{tabular}{cccccc} % five columns, alignment for each
		\hline
		    & $\lambda$ & $Q_\mathrm{ext}~(\upmu$m) & $Q_\mathrm{abs}~(\upmu$m) & $Q_\mathrm{sca}~(\upmu$m) & $g$ \\
		\hline
		DDSCAT & 0.4 & 4.6743 & 2.8698 & 1.8046  & 0.84068 \\
            CORAL  & 0.4 & 4.6551 & 2.8546 & 1.8005  & 0.84126 \\
            \hline
            DDSCAT & 0.8 & 3.5604 & 2.6364 & 0.92414 & 0.74129 \\
            CORAL  & 0.8 & 3.5637 & 2.6424 & 0.92130 & 0.74295 \\
            \hline
            DDSCAT & 1.5 & 2.3904 & 2.0517 & 0.33886 & 0.61466 \\
            CORAL  & 1.5 & 2.3937 & 2.0562 & 0.33749 & 0.61438 \\
            
		\hline
	\end{tabular}
\end{table}

\subsection{Speed tests} \label{appendix:speed_tests}

In Table \ref{table:Computation times}, we list the computation time to produce the data required for the graphs in Fig.~\ref{fig:results_branched}-\ref{fig:results_compact} (an angle average over 252 different angles, two polarisation states, and 100 different wavelengths). These were completed using OpenMP and 28 threads of an AMD EPYC 7702P 64-core Processor. We show the \% error in $C_\mathrm{ext}$ to demonstrate the trade-off between decreased calculation time, but lower accuracy, for low-resolution DDA. It should calculations in \texttt{DDSCAT} and \texttt{OPTOOL} can be faster than \texttt{CORAL}, so this data should only be used as a rough comparison of times for varying dipole numbers rather than a thorough comparison of the methodologies themselves.

\setlength{\tabcolsep}{2pt} % reduce white space to fit table onto one page
\begin{table}
	\centering
	\caption{Computation time for the different models used in this study. The \% err. on the Mie and MMF values vary depending on the shape (see Table~\ref{table:Beta_and_lambda_min_values}), but computation times are similar in each case.}
	\label{table:Computation times}
	\begin{tabular}{crrc} % four columns, alignment for each
		\hline
	    Shape type   & Dipole number $N$ & Computation time (s) & \% err. $C_\mathrm{ext}$ \\
		\hline
		                       &    131 &        117 & 7.6 \% \\
              Linear branched     &  1,085 &        810 & 4.0 \% \\
                                  &  9,410 &     67,612 & 1.6 \% \\
                                  & 81,313 & 19,386,369 &  -  \\
            \cmidrule{2-4}
                                  &      29 &         49 & 3.2 \% \\
              Elongated cluster   &     233 &        101 & 1.0 \% \\
                                  &   1,787 &      2,357 & 2.0 \% \\
                                  &  15,604 &    173,754 & 0.9 \% \\
                                  & 131,130 & 37,317,262 &  -  \\
            \cmidrule{2-4}
                                  &      90 &        93 & 3.2 \% \\
              Compact cluster     &     675 &       395 & 1.0 \% \\
                                  &   5,043 &    18,208 & 2.0 \% \\
                                  &  43,730 & 2,023,438 &  -  \% \\
		\hline
                                  &    Mie &      1 & between 25-37 \\
                                  &    MMF &    729  & between 19-26 \\
            \hline
	\end{tabular}
\end{table}

\section{Residuals} \label{appendix:residuals}

\begin{figure*}
\centering
\begin{subfigure}[t]{0.785\textwidth}
\centering
\includegraphics[width=\textwidth]{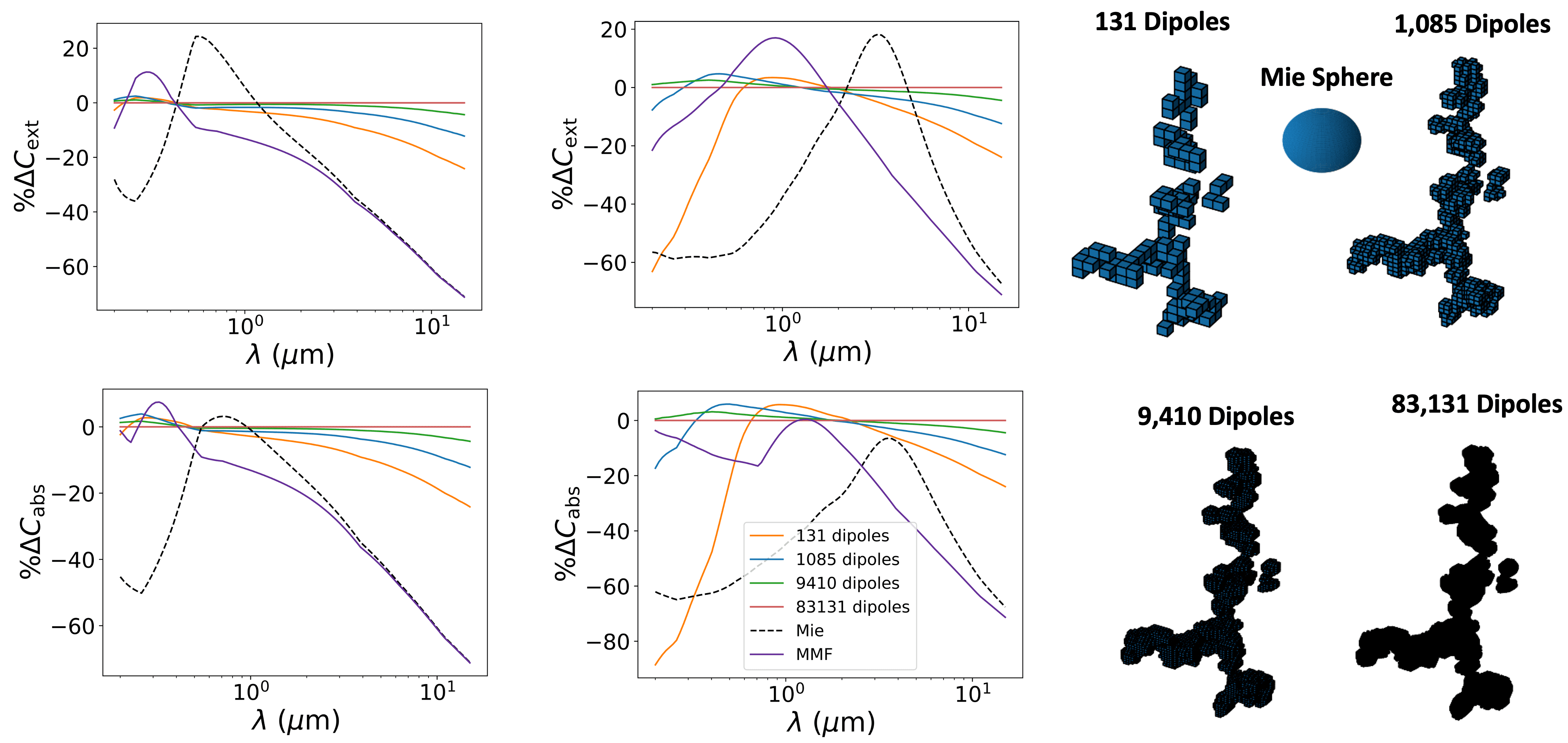} 
\caption{Residuals for the Linear branched fractal in Fig.~\ref{fig:results_branched} (in the same arrangement as the original figure, with the left two graphs representing particles with an equivalent radius ($R_\mathrm{mie}$) of 0.1~$\upmu$m, and the right two graphs representing an equivalent radius of 0.5~$\upmu$m.} \label{fig:timing1}
\end{subfigure}

\begin{subfigure}[t]{0.785\textwidth}
\centering
\includegraphics[width=\textwidth]{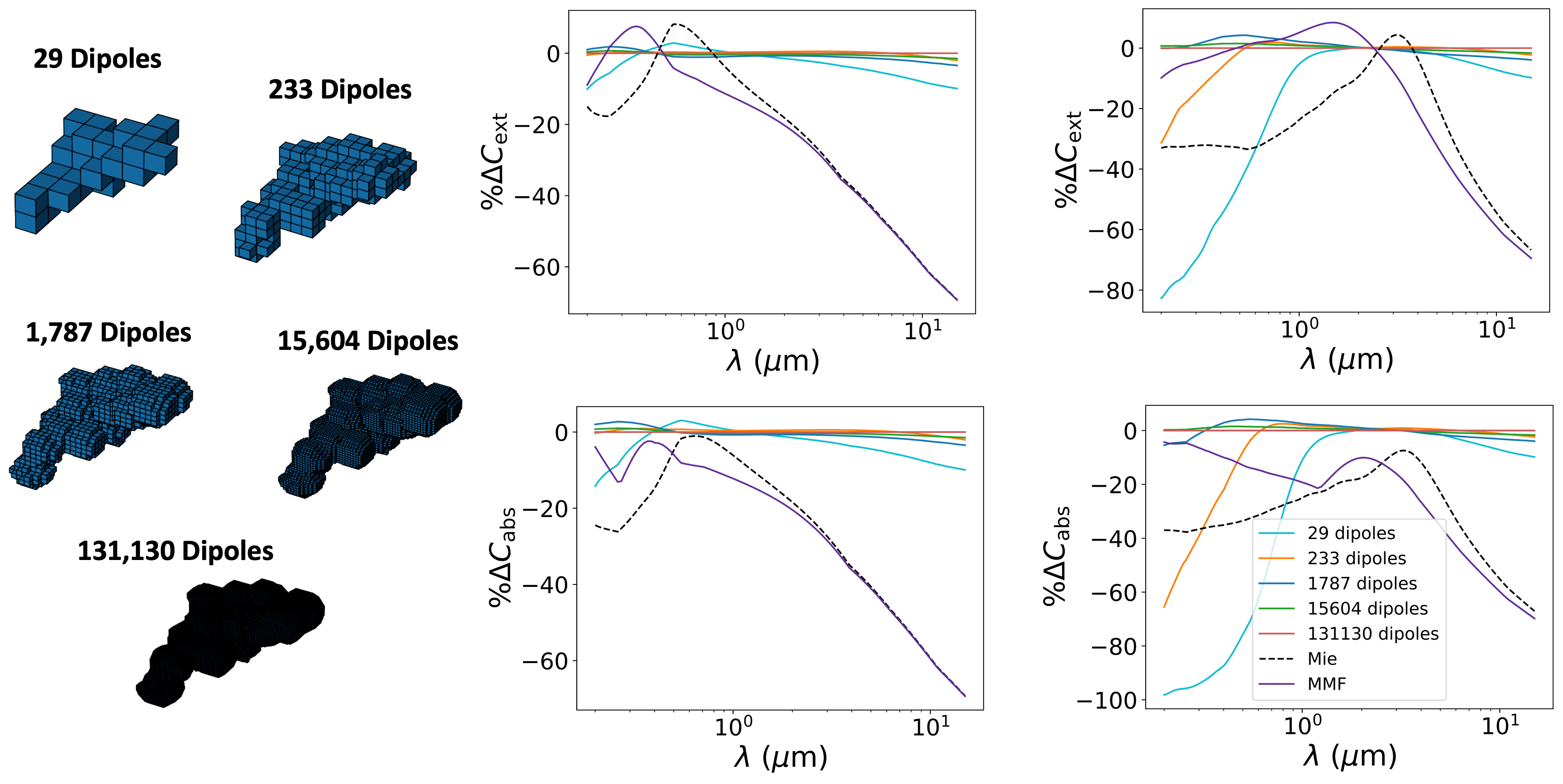} 
\caption{As above but for the elongated cluster in Fig.~\ref{fig:results_elongated}.} \label{fig:timing2}
\end{subfigure}

\begin{subfigure}[t]{0.785\textwidth}
\centering
\includegraphics[width=\textwidth]{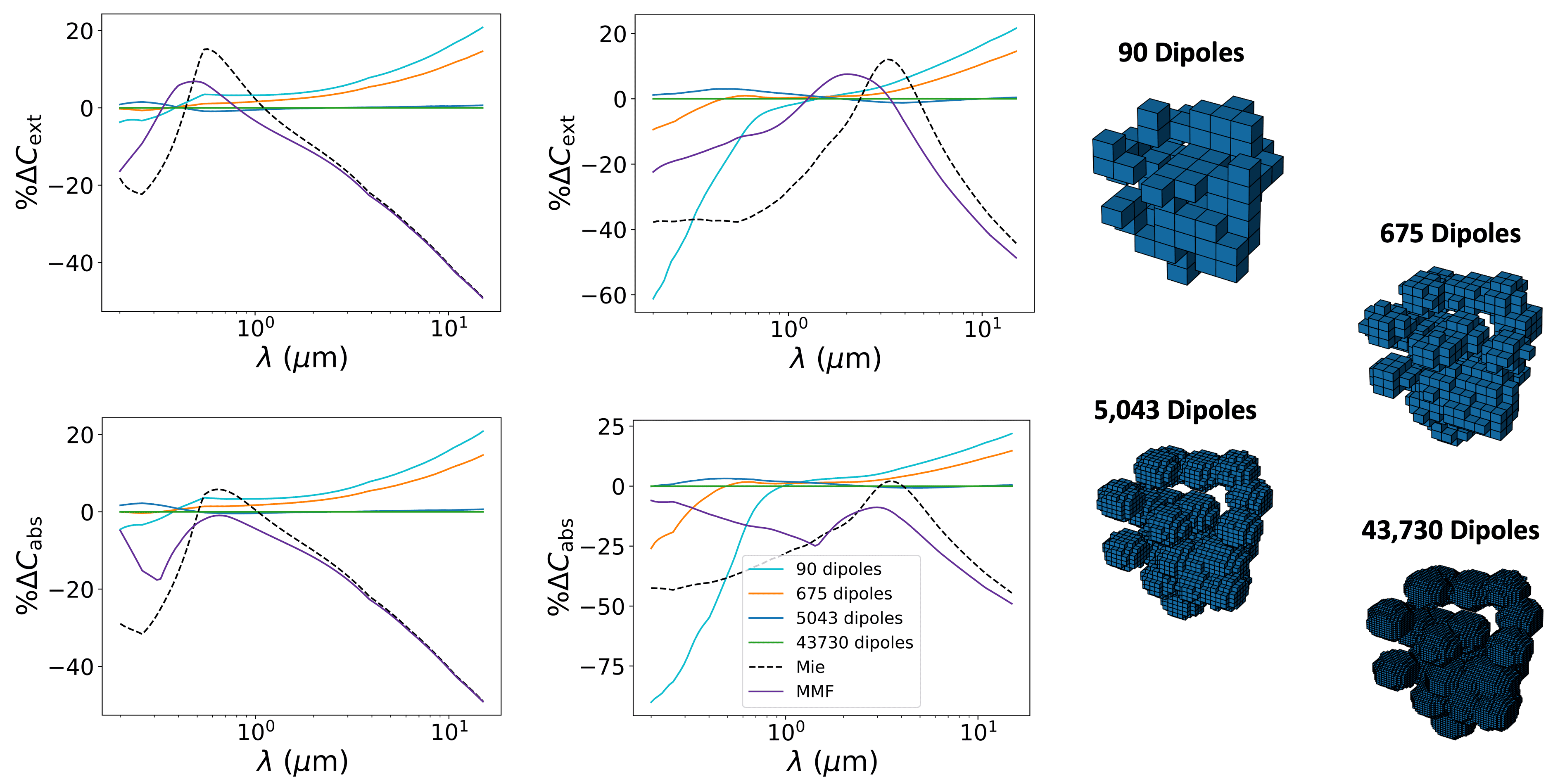} 
\caption{As above but for the compact cluster in Fig.~\ref{fig:results_compact}.} \label{fig:timing3}
 \end{subfigure}

 \caption{Residual values versus the benchmark model for each of the three shape types in in Fig.~\ref{fig:results_branched}-\ref{fig:results_compact}.}
\label{fig:residuals_main}
\end{figure*}

\begin{figure*}
    \includegraphics[width=\textwidth]{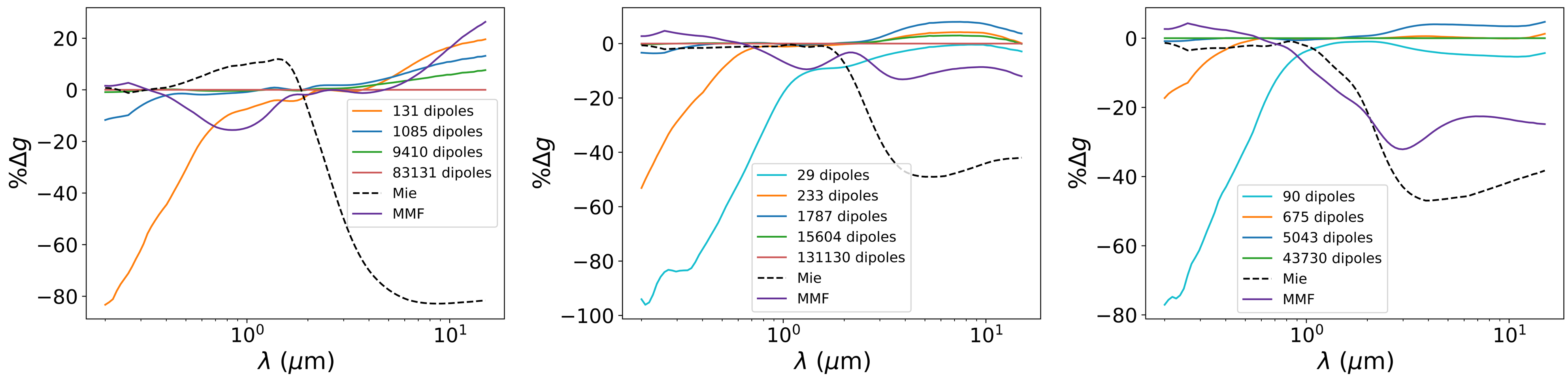}
    \caption{Residuals for the asymmetry parameter $g=\braket{\cos\theta}$ in Fig.~\ref{fig:results_g_asymmetry}), plotted for each resolution of the linear branched, elongated cluster, and compact cluster shape types (from left to right).}
    \label{fig:residuals_g_asymmetry}
\end{figure*}

Fig.~\ref{fig:residuals_main} shows the residuals versus the most accurate benchmark model for the values of $C_\mathrm{ext}$ and $C_\mathrm{abs}$ for each shape type, for all of the graphs in Figs.~\ref{fig:results_branched}-\ref{fig:results_compact}. We also show the residuals for asymmetry parameter $g$ (original data from Fig.~\ref{fig:results_g_asymmetry}) in Fig.~\ref{fig:residuals_g_asymmetry}.

%%%%%%%%%%%%%%%%%%%%%%%%%%%%%%%%%%%%%%%%%%%%%%%%%%

% Don't change these lines
\bsp	% typesetting comment
\label{lastpage}
\end{document}